\documentclass[11pt]{article}
\pdfoutput=1
\usepackage{amsmath}
\usepackage{amssymb}
\usepackage{graphicx,bbm,mathrsfs}
\usepackage{nicefrac}
\usepackage{slashed}
\usepackage{bbm}
\usepackage{geometry}
\usepackage{empheq}
\usepackage{stackrel}
\usepackage{ulem}
\usepackage{jheppub}     
\usepackage{setspace}
\usepackage{relsize}
\usepackage{empheq}
\usepackage{pifont}
\usepackage{mathtools}
\usepackage{enumitem}
\usepackage{dcolumn}   
\usepackage{bm}        
\usepackage{graphicx,mathrsfs}
\usepackage{nicefrac}
\usepackage{multirow}
\usepackage{color}
\usepackage{mathtools}
\usepackage{makecell}
\usepackage{environ} 
\usepackage{lipsum} 
\usepackage{wasysym}
\usepackage{hhline,colortbl}  
\usepackage{tikz}
\usepackage{graphicx}
\usepackage{ulem}
\usepackage{mdframed}
\usepackage{xcolor}
\definecolor{EqFrame}{RGB}{235,245 ,250 }

\newcommand*\annotatedFigureBoxCustom[8]{\draw[] ;\node at (#4) [] {\textbf{#3}};}
\newcommand*\annotatedFigureBox[4]{\annotatedFigureBoxCustom{#1}{#2}{#3}{#4}{black}{black}{black}{black}}

\newenvironment {annotatedFigure}[1]{\centering\begin{tikzpicture}
    \node[anchor=south west,inner sep=0] (image) at (0,0) { #1};\begin{scope}[x={(image.south east)},y={(image.north west)}]}{\end{scope}\end{tikzpicture}}

 \NewEnviron{Smaller11}{
           \scalebox{1.1}{$\BODY$} 
 } 
 \NewEnviron{Smaller08}{
           \scalebox{0.8}{$\BODY$} 
 }

\newcommand{\cmark}{\ding{51}}%
\newcommand{\xmark}{\ding{55}}%

\newcommand{\be}{\begin{equation}}
\newcommand{\ee}{\end{equation}}
\newcommand{\bea}{\begin{eqnarray}}
\newcommand{\eea}{\end{eqnarray}}

\newcommand{\M}{\cal M}
\newcommand{\Mnu}{{\cal M}_\nu}
\newcommand{\MnuM}{{\cal M}_\nu^-}
\newcommand{\MnuP}{{\cal M}_\nu^+}

\newcommand{\s}{\rm s}
\renewcommand{\u}{\Delta}

\newcommand{\f}{h}
\newcommand{\g}{k}
\newcommand{\fund}{ \rm f}

\usepackage{xcolor}

\renewcommand{\S}{{\cal S}}

\newcommand{\nn}{\nonumber}

\renewcommand{\L}{L}

\usepackage{titlesec}

\titleformat*{\section}{\Large\bfseries}
\titleformat*{\subsection}{\large\bfseries}
\titleformat*{\subsubsection}{\large\bfseries}
\titleformat*{\paragraph}{\large\bfseries}
\titleformat*{\subparagraph}{\large\bfseries}

\makeatletter
\newcommand*{\prodsym}{%
  \DOTSB
  \mathop{
    \mathchoice
      {\rlap{\kern.3em\rotatebox[origin=c]{-90}{}}{\prod}}
      {\vcenter{\rlap{\kern.2em\rotatebox[origin=c]{-90}{}}}{\prod}}
      {\sum}{\sum}
  }\slimits@
}
\makeatother

\makeatletter
\DeclareFontFamily{OMX}{MnSymbolE}{}
\DeclareSymbolFont{MnLargeSymbols}{OMX}{MnSymbolE}{m}{n}
\SetSymbolFont{MnLargeSymbols}{bold}{OMX}{MnSymbolE}{b}{n}
\DeclareFontShape{OMX}{MnSymbolE}{m}{n}{
    <-6>  MnSymbolE5
   <6-7>  MnSymbolE6
   <7-8>  MnSymbolE7
   <8-9>  MnSymbolE8
   <9-10> MnSymbolE9
  <10-12> MnSymbolE10
  <12->   MnSymbolE12
}{}
\DeclareFontShape{OMX}{MnSymbolE}{b}{n}{
    <-6>  MnSymbolE-Bold5
   <6-7>  MnSymbolE-Bold6
   <7-8>  MnSymbolE-Bold7
   <8-9>  MnSymbolE-Bold8
   <9-10> MnSymbolE-Bold9
  <10-12> MnSymbolE-Bold10
  <12->   MnSymbolE-Bold12
}{}

\let\llangle\@undefined
\let\rrangle\@undefined
\DeclareMathDelimiter{\llangle}{\mathopen}%
                     {MnLargeSymbols}{'164}{MnLargeSymbols}{'164}
\DeclareMathDelimiter{\rrangle}{\mathclose}%
                     {MnLargeSymbols}{'171}{MnLargeSymbols}{'171}
\makeatother

\begin{document}

\vspace*{4mm}

\thispagestyle{empty}

\begin{center}

\begin{minipage}{20cm}
\begin{center}
\hspace{-5cm }
\LARGE
\sc
Entanglement Entropy and Thermal  Phase Transitions
 \\
\hspace{-5cm }    from Curvature Singularities
\end{center}
\end{minipage}
\\[30mm]

\renewcommand{\thefootnote}{\fnsymbol{footnote}}

{\large  
Sergio~Barbosa$^{\, a}$ \footnote{sergio.barbosa@aluno.ufabc.edu.br}\,, 
Sylvain~Fichet$^{\, a}$ \footnote{sylvain.fichet@gmail.com}\,, 
Eugenio~Meg\'{\i}as$^{\, b}$ \footnote{emegias@ugr.es}\,,
Mariano~Quir\'os$^{\, c}$ \footnote{quiros@ifae.es}\,
}\\[12mm]
\end{center} 
\noindent

${}^a\!$ 
\textit{CCNH, Universidade Federal do ABC,} 
\textit{Santo Andre, 09210-580 SP, Brazil}

${}^b\!$ 
\textit{Departamento de F\'{\i}sica At\'omica, Molecular y Nuclear and} \\
\indent \; \textit{Instituto Carlos I de F\'{\i}sica Te\'orica y Computacional,} \\
\indent \; \textit{Universidad de Granada, Avenida de Fuente Nueva s/n, 18071 Granada, Spain}

${}^c\!$  
\textit{Institut de F\'{\i}sica d'Altes Energies (IFAE) and} \\
\indent \; \textit{The Barcelona Institute of  Science and Technology (BIST),} \\
\indent \; \textit{Campus UAB, 08193 Bellaterra, Barcelona, Spain}

\addtocounter{footnote}{-4}

\vspace*{8mm}
 
\begin{center}
{  \bf  Abstract }
\end{center}
\begin{minipage}{15cm}
\setstretch{0.95}

  We study holographic entanglement entropy and revisit thermodynamics and confinement in the dilaton-gravity system. Our analysis focuses   on  a solvable class of backgrounds that includes AdS and linear dilaton spacetimes as particular cases, with some results  extended   to  general warped metrics. A general lesson is that the behavior of the holographic theory is tied to the bulk curvature singularities.
We  find that a singular background is confining if and only if  \textit{i)} the singularity coincides with  a  boundary or \textit{ii)} it is the linear dilaton.  
In the  former case, for which the singularity cuts off spacetime, we  demonstrate that both entanglement entropy and thermodynamics exhibit a first order phase transition.
In the linear dilaton case we find instead that both  entanglement entropy and thermal phase transitions are   of second order. Additionally, along the process we thoroughly derive the  radion effective action at quadratic order.

    \vspace{0.5cm}
\end{minipage}

\newpage
\setcounter{tocdepth}{2}
\tableofcontents
\newpage

\section{Introduction \label{se:intro}}

Some classes of holographic spacetimes beyond pure Anti-de Sitter  feature a curvature singularity. 
 The singularity may be either at finite or infinite conformal distance from 
 the boundary on which the lower-dimensional dual theory is computed. That is, we have schematically \\
\vspace{-0.5cm}
\begin{center}
\includegraphics[width=0.35\linewidth,trim={2cm 3.1cm 0cm 4.25cm},clip]{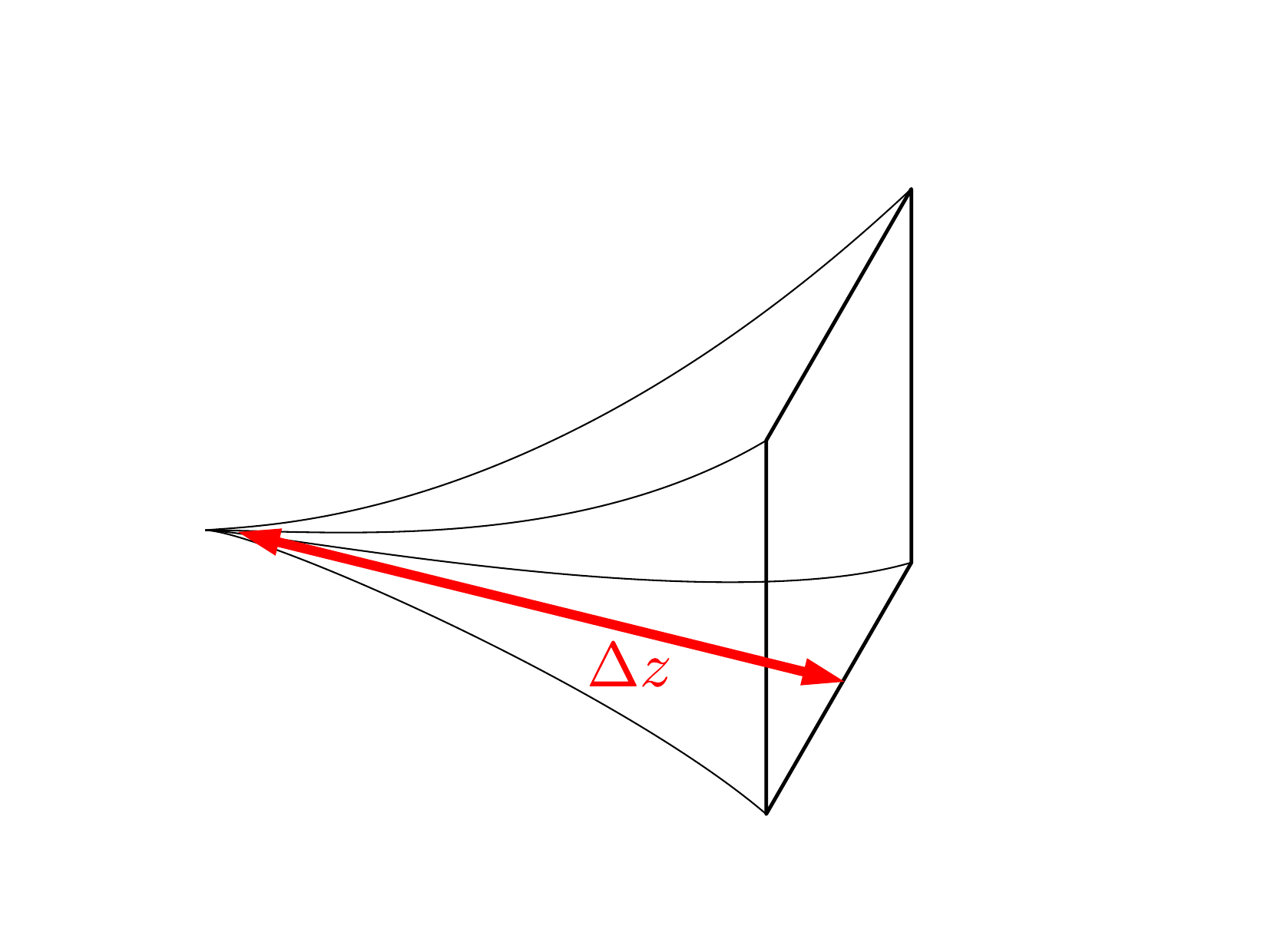}
 \end{center}
with $\Delta z$ finite  or not. In this paper we argue that this simple distinction  controls key properties of the dual theory,
 including the behavior of entanglement entropy, thermal phase transitions, and confinement.

The phenomenon of confinement remains an unsolved puzzle of QFT at strong coupling. One route to understand confinement is to describe it holographically, using a gauge-gravity duality that may potentially encode the strong dynamics in a gravitational higher-dimensional spacetime. While the most precise and well-established duality of this kind is the AdS/CFT correspondence~\cite{Aharony:1999ti, Zaffaroni:2000vh, Kap:lecture,Nastase:2007kj}, CFTs  cannot confine since they have no scale.  
 {The holographic view of confinement, if it exists, must come from a different, less symmetric spacetime beyond AdS, that allows for the emergence of a confinement scale. }

The gauge-gravity correspondence being an offspring of string theory, natural candidates for holography beyond AdS are motivated by non-critical string-inspired models~\,\cite{Polchinski:2001tt,Erlich:2005qh,DaRold:2005mxj,Gursoy:2007cb, Gursoy:2007er}.  The low-energy effective action in such models describes spacetime augmented with an extra scalar field, the dilaton.  Going beyond AdS/CFT naturally  leads  to exploring the properties of the dilaton-gravity system.

One reason to expect that dilatonic backgrounds are holographic is the existence of a known example: the linear dilaton spacetime, whose corresponding dual theory is little string theory (LST).\,\footnote{
See \cite{Aharony:1999ks,Kutasov:2001uf, Antoniadis:2021ilm} for LST reviews, \cite{Seiberg:1997zk,Berkooz:1997cq,
Giveon:2017nie,
Giribet:2017imm, Asrat:2017tzd, Araujo:2018rho,Chakraborty:2020fpt,Chakraborty:2020yka,Georgescu:2022iyx, Guica_lecture, Chang:2023kkq,Chakraborty:2023mzc,Chakraborty:2023zdd,Aharony:2023dod, Fichet:2023xbu} for formal holographic aspects of the linear dilaton background, and \cite{Antoniadis:2001sw, Antoniadis:2011qw, Csaki:2018kxb,Megias:2019vdb,Megias:2021mgj,Megias:2021arn,Antoniadis:2021ilm, Fichet:2022ixi, Fichet:2022xol} for field theoretical  studies of the LD background and related  phenomenological developments.} 
More generally it is widely accepted that a (less predictive) holographic principle formulated in terms of information content applies to any spacetime~\cite{tHooft:1993dmi,Susskind:1994vu,tHooft:1999rgb,Bousso:1999xy,Bousso:2002ju}. These are strong motivations to investigate the holography of the dilaton-gravity system, including the holographic version of  confinement.\,\footnote{See e.g. \cite{Karch:2006pv,Gursoy:2007cb,Gursoy:2007er, Batell:2008zm, Colangelo:2008us, Cabrer:2009we,vonGersdorff:2010ht,Megias:2017czr, Faedo:2024zib} for related studies of the dilaton-gravity system. }

Hints that certain non-AdS spacetime backgrounds describe confining dynamics have been gradually gathered, using holographic descriptions of the quark-antiquark potential~\cite{Witten:1997ep,Hanany:1997hr,Maldacena:1998im,Kinar:1998pj,Polyakov:1998ju,Rey:1998ik, Andreev:2006ct, Andreev:2006nw, Gursoy:2007er,Galow:2009kw}, of thermal phase transitions~\cite{Horigome:2006xu,Gursoy:2008za,Konstandin:2010cd,Megias:2010ku,Megias:2018sxv,Megias:2020vek,Megias:2023kiy,Mishra:2024ehr}, and of entanglement entropy \cite{Klebanov:2007ws,Bah:2007kcs,Fujita:2008zv,Engelhardt:2013jda,
Chakraborty:2018kpr,Fujita:2020qvp,daRocha:2021xwq,Jokela:2023lvr,
Kol:2014nqa,Jokela:2020wgs,Grieninger:2023pyb,Fatemiabhari:2024aua}. Such phenomena can be viewed as probes that test whether a background is confining, based on the broad qualitative features expected from confining theories.

While a number of studies of  holographic entanglement entropy --- as a probe of confinement --- have been done in stringy and supersymmetric backgrounds (see~\cite{Klebanov:2007ws,Bah:2007kcs,Fujita:2008zv, Abajo-Arrastia:2010ajo, Engelhardt:2013jda,Bagchi:2014iea, Erdmenger:2017gdk,Chakraborty:2018kpr,Fujita:2020qvp, Casalderrey-Solana:2020vls, daRocha:2021xwq,Jokela:2023lvr,
Kol:2014nqa,Jokela:2020wgs,Fatemiabhari:2024aua}),  holographic entanglement has not been studied in the simple  dilaton-gravity system, to the best of our knowledge, apart from a brief case study~\cite{Li:2024lrh}.  One  goal of this paper is to fill this  gap.

Another objective of this study is to elucidate the relationship between curvature singularities and confinement. These singularities, induced by the backreaction of the metric on the dilaton vacuum expectation value (vev), are prevalent in dilaton-gravity models. Here, we will categorize the singularities in terms of conformal distance to the brane, a  distinction that we will show to be linked to the confining nature of the background.

A third objective of this paper is to provide  a unified view and a simple set of intuitions 
for the various probes of confinement, that can be applied to e.g.~asymptotically AdS backgrounds. Here we compute all these phenomena in a  class of non-asymptotically AdS backgrounds ({previously used in \cite{Fichet:2023dju}, see also \cite{Gursoy:2015nza,Betzios:2017dol,Betzios:2018kwn,Gursoy:2021vpu}})
 that retains the essence of the features we want to highlight. This class of backgrounds has a single continuous parameter and
contains both the linear dilaton and AdS as particular cases.

Our study is structured as follows. 
In section \ref{se:setup} we define the {$D$-dimensional} dilaton-gravity system of our interest, 
and classify its  curvature singularities. 
In section~\ref{se:string} we analyze the string-based notion of confinement in this background, developing intuitions that also apply to asymptotically-AdS backgrounds. 
In section~\ref{se:EE} we compute the holographic entanglement entropy of a $(D-2)$-strip localized on the boundary, including finite temperature corrections.  
In section \ref{sec:radion} we compute the 
radion effective action. At zero temperature this controls the stability of the brane-dilaton system. At finite temperature the radion effective action gives access to the free energy. We explore the structure of phase transitions in section \ref{sec:PT}. 
Our findings are summarized in section \ref{se:conclusion}.  Appendix \ref{app:aprimelimit} contains a derivation of  a  property used in the analysis of metric zeros.

\section{Dilaton-Gravity  and Singularities}

\label{se:setup}

We consider a spacetime with $D=d+1$ dimensions.
The action of the $D$-dimensional dilaton-gravity system {in the Einstein frame} is
\begin{eqnarray}
\S &=& \int d^Dx  \sqrt{g}  \bigg(  \frac{M_D^{D-2}}{2} {}^{(D)}R  - \frac{1}{2} (\partial_M \phi)^2 - V(\phi)  \bigg) \nonumber \\ 
&&-\int_{\textrm{brane}} d^dx \sqrt{\bar g} \left(V_b(\phi_b) + \Lambda_b - M_D^{D-2} K \right) ~ + ~ \S_{\textrm{matter}}  
\,, \label{eq:action}
\end{eqnarray}
where ${}^{(D)}R$ is the  scalar curvature, $\phi$ the dilaton field, $M_D$ the fundamental $D$-dimensional Planck scale, {and $g_{MN}$ the bulk metric with mostly plus signature}. 
$\S_{\textrm{matter}}$ encodes the quantum fields living on this background. 

The spacetime supports a $(d-1)$-brane, with induced metric  $\bar g_{\mu\nu}$, {tension $\Lambda_b$ and a localized potential $V_b(\phi_b)$,  where $\phi_b \equiv \phi|_{\rm brane}$ stands for the value of the dilaton field at the brane. $K$ is the extrinsic curvature that appears in  the Gibbons-Hawking-York (GHY) boundary term.} 
The $V_b(\phi_b)$ potential stabilizes  $\phi_b$ to the vacuum expectation value (vev) $\langle\phi_b\rangle \equiv  v_b$, which  determines completely the background. 

The $V_b(\phi_b)$ potential  is assumed to satisfy a mild condition, specified further below,  that ensures stability of spacetime \cite{Fichet:2023dju}.    
This assumption and the fact that $V_b(\phi_b)$  stabilizes $\phi_b$ are sufficient in the scope of our study,  hence the explicit form of  $V_b(\phi_b)$ does not need to be further specified.
 We   use the convention $V_b(v_b)=0$ without loss of generality.

The most general metric ansatz we consider is the 
warped foliation
\be
ds^2 =g_{MN}dx^Mdx^N = g_{xx}(r)\eta_{\mu\nu}dx^\mu dx^\nu  +g_{rr}(r) dr^2\,, \label{eq:ds2_gen}
\ee
with arbitrary $g_{xx}$, $g_{rr}$ coefficients. 
Some of our results will be derived with this general metric, using sometimes monotonicity assumptions involving the metric coefficients.

\subsection{Branes and Holography}

The stabilization of the dilaton-gravity system involves a brane. We will see throughout this paper that the dual holographic theory is naturally defined on  this brane.

\subsubsection{Branes}

From the low-energy viewpoint, a brane is simply an infinitely thin hypersurface living in the higher dimensional spacetime and on which operators and degrees of freedom can be localized \cite{Csaki:2004ay, Sundrum:1998sj,Sundrum:1998ns}.\,\footnote{From the UV viewpoint branes may be viewed as solitons. Similar objects naturally appear in string theory  as D-branes, which are dynamical objects with quantum properties \cite{Polchinski:1996na,Bachas:1998rg}. Black brane solutions also arise in the supergravity limit of string theories \cite{Aharony:1999ti, Duff:1996zn}. } 

All fields, including the graviton, satisfy  boundary conditions on the brane. 
At low-energies, the brane may be thought as a defect with spacetime existing on both sides. In such a case, the dynamics of the brane can be obtained using the Codacci~equation \cite{Shiromizu:1999wj} or equivalently  Israel's junction condition~\cite{Israel:1966rt}. These relate the extrinsic curvatures on both sides of the brane to the brane-localized stress tensor, $\tau_{\mu\nu}$~\cite{Fichet:2023xbu}. One has 
\begin{equation}
[K_{\mu\nu}] \equiv K^+_{\mu\nu} - K^-_{\mu\nu} = - \frac{1}{M_D^{D-1}} \left( S_{\mu\nu}  -  \frac{1}{D-2}  \bar g_{\mu\nu} S \right) \,, \label{eq:Junction}
\end{equation}
where $S_{\mu\nu} = - \Lambda_b \bar g_{\mu\nu} + \tau_{\mu\nu}$. 
A particular case is when both sides are simply the mirror of each other, i.e.~related by a $ Z_2$ symmetry, so that $K_{\mu\nu}^+ = - K_{\mu\nu}^-$ {(as by definition $K_{\mu\nu}$ is $Z_2$-odd)} and then $K_{\mu\nu}^\pm = \pm \frac{1}{2} [K_{\mu\nu}]$. This is the ``$ Z_2$-orbifold'' {picture}
used in braneworld computations such as~\cite{Shiromizu:1999wj}. 

Equivalently, the brane may be thought of as a wall beyond which spacetime ends {instead of being ``mirrored''}. This is called an end-of-the-world (EOW) brane.    In this case, the dynamics of the brane is determined by letting the metric be dynamical on the brane. This produces a Neumann-type boundary condition that takes precisely the form of \eqref{eq:Junction} with zero $K_{\mu\nu}$ on the empty side.  This means that for the $\mathcal M^-_\nu$ space that will be introduced in Sec.~\ref{subsec:Mnu_spacetime}, one has $K^+_{\mu\nu} = 0$ so that   $K_{\mu\nu}^- = - [K_{\mu\nu}]$ with differs by a factor of 2 from the result within the $ Z_2$-orbifold convention.~\footnote{For the $\mathcal M_\nu^+$ space, one would have instead $K^-_{\mu\nu} = 0$ and $K_{\mu\nu}^+ = [K_{\mu\nu}]$.} The EOW picture is equivalent to the $ Z_2$-orbifold picture, and  one can be translated into the other just paying attention to some factors of $2$ related to the $Z_2$ mirroring. In this paper we use the $ Z_2$-orbifold  picture, which is needed in sections \ref{sec:radion} and \ref{sec:PT}.

\subsubsection{Holography on the Brane}

 The holographic theory is defined by integrating out the bulk degrees of freedom, providing  an effective $d$-dimensional theory supported on the brane. 
In the context of  dilaton gravity, the existence of the brane is mandatory in the sense that it stabilizes the dilaton-gravity system and fixes the physical scale. 
The brane can optionally be  thought as a regulator and be sent to a conformal boundary to recover the asymptotic approach to holography. 

The advantage of our approach is that well-understood braneworld-like computation techniques such as those from  \cite{Shiromizu:1999wj, Langlois:2002ke,Langlois:2003zb} can be applied.
Non-trivial consistency checks of our solutions made within this formalism can be found in  \cite{Fichet:2022ixi,Fichet:2023xbu,Fichet:2023dju}.  

\subsection{Solving the Brane-Dilaton-Gravity System }
\label{subsec:sol_BDG}

The bulk field equations are 
\begin{eqnarray}
  0 &=& {}^{(D)}R_{MN} - \frac{1}{2} g_{MN} {}^{(D)}R - \frac{1}{M_D^{D-2}} \, \partial_M \phi  \partial_N \phi + \frac{1}{2 M_D^{D-2}} g_{MN} (\partial_A\phi)^2 + \frac{1}{M_D^{D-2}} \, g_{MN}  V(\phi)  \,, \nonumber \\ \label{eq:Eg}\\
  0 &=&  \frac{1}{\sqrt{g}} \partial_M\left(\sqrt{g} g^{MN}\partial_N \phi \right) - \frac{\partial V}{\partial \phi} \,. \label{eq:Ephi}
  \end{eqnarray}
In addition, the brane-localized terms in the action induce some boundary conditions at the brane, that are of the form $[K_{\mu\nu}] \equiv K_{\mu\nu}^+ - K_{\mu\nu}^- = \mathcal F(\bar g_{\mu\nu}, \phi_b)$ and $[\phi^\prime] \equiv \phi^\prime_+ - \phi^\prime_- = \mathcal G(\bar g_{\mu\nu}, \phi_b)$ (see Eq.~(\ref{eq:Junction}))~\cite{Csaki:2000zn}. The solutions to the field equations have integration constants that need careful analysis: some are gauge redundancies, other are physically meaningful.  
Some integration constants are fixed by the boundary conditions at the brane. On the other hand, a combination of integration constants is fixed by the requirement that the vev does not change with the brane location,  $\frac{\partial v_b}{\partial r_b}=0$ \cite{Fichet:2023xbu}, this being a natural consequence of the fact that the $V_b(\phi_b)$ potential is independent on the brane location.  We refer to \cite{Fichet:2023xbu} for a detailed discussion.

\subsection{The $\Mnu$ Spacetime}
\label{subsec:Mnu_spacetime}

We introduce the reduced bulk potential $V(\phi) \equiv (D-2)
M_D^{D-2} \bar V(\bar\phi)$ and the reduced dilaton field $\phi \equiv
\sqrt{(D-2) M_D^{D-2}} \bar \phi $. $\bar V$ {has mass dimension 2},
while $\bar\phi$ is dimensionless.  The $\Mnu$ spacetime used in this
work is defined by setting the reduced bulk potential to
\begin{equation}
\bar V(\bar\phi) = - \frac{1}{2}(D-1-\nu^2) k^2 e^{2 \nu \bar\phi} \,. \label{eq:barV}
\end{equation}
Here  $\nu$ is a real parameter that we take positive without loss of generality.

{We assume that the brane is flat, i.e.~$r_b$ does not depend on the space coordinates.  While the brane could evolve in time, as studied in \cite{Fichet:2023xbu,Fichet:2023dju}, see e.g.~\cite{Hebecker:2001nv,Langlois:2002ke,Langlois:2003zb} for AdS,   here we consider a static spacetime where $r_b$ is constant in $x^\mu$. }

In the absence of a black hole horizon, the solutions to the field equations following from the action \eqref{eq:action} are found to be \cite{Fichet:2023dju}
\begin{eqnarray}
ds^2_{\nu,r_b} &=& \left(\frac{ r}{L}\right)^2 \eta_{\mu\nu}dx^\mu dx^\nu + 
\left(\frac{  r}{r_b}\right)^{2\nu^2}
\frac{1}{(\eta r)^2}
d r^2\,, \label{eq:ds2_Mnu} \\
\bar\phi(r) &=& \bar \phi_b - \nu \log\left( \frac{r}{r_b} \right)  \,, \label{eq:barphi_r}
\end{eqnarray}
with $r\in\mathbb{R}_+$. {$L$ is a constant with dimension of length}. 
In \eqref{eq:barphi_r}, $\bar\phi_b = \frac{1}{\sqrt{(D-2)M_D^{D-2}}} \phi_b$ is the value of the reduced dilaton field on the brane. The dilaton vev and the other parameters of the metric combine to form the physical scale $\eta \equiv k \, e^{\nu \bar v_b}$, that appears in observable quantities.  

The brane at $r=r_b$ partitions $\M_\nu$ into two regions
\be
\MnuM=\Mnu \big|_{r\in(0,r_b]} \,,\quad\quad \MnuP=\Mnu \big|_{r\in[r_b,\infty)} \,. 
\label{eq:Mnu_halves}
\ee
Our focus in this work is the $\MnuM$ space. We  mention sometimes $\MnuP$ for comparison.

\begin{figure}[t]
\centering
\includegraphics[trim={0cm 2cm 0cm 3cm},clip,width=0.8\textwidth]{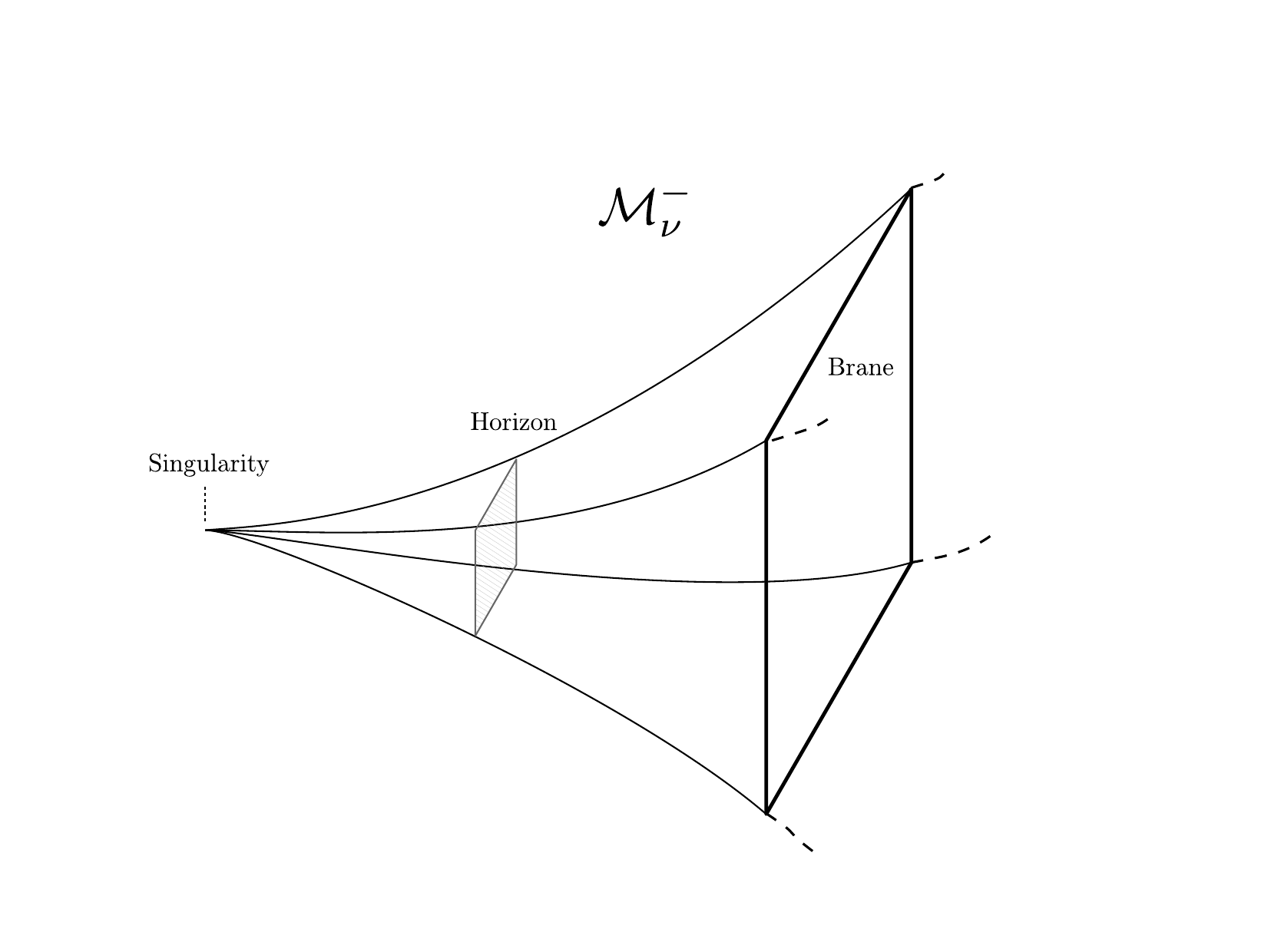}
\caption{\it   Sketch of the  $\MnuM$ spacetime. The curvature singularity can be hidden behind a horizon for $\nu<\sqrt{D-1}$. 
}
\end{figure}

\subsubsection*{Conformal frame}

We sometimes use  conformal coordinates throughout this work.  For any $\nu\neq1$ these are defined as
  {
  \be
   z=\frac{L}{\eta\, r_b} \frac{1}{|\nu^2-1|}\left(\frac{r}{r_b}\right)^{\nu^2-1}\,, \qquad \nu\neq 1 \,,
  \ee
  }
with the domain $z\in \mathbb{R}_+$. 
The metric in conformal coordinates  reads
  {
  \be
  ds^2_{\nu,r_b} = \left(\frac{r_b}{L}\right)^{\frac{2\nu^2}{\nu^2-1}}  \left(|\nu^2-1|\eta z\right)^{\frac{2}{\nu^2-1}}
  (\eta_{\mu\nu}dx^\mu dx^\mu  +dz^2)\,.
  \ee
  }
In these coordinates, the $\MnuM$ and $\MnuP$ regions defined in \eqref{eq:Mnu_halves} are given by 
\begin{align}
\MnuP&=\Mnu \big|_{z\in(0,z_b]} \,,\quad \MnuM=\Mnu \big|_{z\in[z_b,\infty)} \,\quad {\rm if}\quad \nu < 1   \,,
\nonumber \\
\MnuM &=\Mnu \big|_{z\in(0,z_b]} \,,\quad \MnuP = \Mnu \big|_{z\in[z_b,\infty)} \,\quad {\rm if}\quad \nu > 1 \,,
\label{eq:M_cases}
\end{align}
where
\be
z_b=\frac{L}{\eta\, r_b} \frac{1}{|\nu^2-1|}\,.
\ee

The special case $\nu=1$ is the linear dilaton spacetime, for which 
\be z= \pm \frac{L}{r_b\eta}\log\frac{r}{L} \label{eq:zLD} \,.\ee 
Importantly, the domain in this case is $z\in \mathbb{R}$.  The freedom of sign in \eqref{eq:zLD}  is reminiscent of the conformal inversion symmetry of the LD spacetime  \cite{Fichet:2023xbu}.

\subsection{The  Black Hole}

Extending the result from \cite{Fichet:2023dju},
the $\MnuM$ spacetime admits a planar black hole solution:\,\footnote{The black hole in the 4D linear dilaton case was independently found in \cite{Karananas:2020fwx}. }
\be
ds^2_{\nu,r_b} = \left(\frac{ r}{L}\right)^2 \left( -f(r) d\tau^2 + d{\bf x}^2  \right) +  \frac{1}{f(r)}
\left(\frac{  r}{r_b}\right)^{2\nu^2}
\frac{1}{(\eta r)^2} d r^2   \,, \label{eq:ds2_Mnu_BH}
\ee
with {the blackening factor}
\begin{equation}
f(r) = 1 - \left( \frac{r_h}{r} \right)^{D-1-\nu^2} \,,
\end{equation}
{where $r_h$ is the location of the horizon.}

\subsection{{On} Singularities and Boundaries}

\label{se:singularities}

For clarity, throughout this work,  we refer to a boundary  as \textit{a regular} boundary, to emphasize the distinction with the notion of conformal boundary. In the vicinity of a regular boundary,  spacetime is isomorphic to the half flat space. A conformal boundary is similarly defined, but  up to a suitable Weyl rescaling to flat space.

\subsubsection{Metric Zeros and Singularities}

\label{se:metric_zeros}

We show a condition under which the vanishing of  the whole metric  implies divergence of the curvature. 

We go to conformal coordinates denoted $(x^\mu, z)$ for which the coefficients of the general metric \eqref{eq:ds2_gen} are $g_{xx}(z)= g_{zz}(z) \equiv a(z)$. 
We refer to a zero of the warp factor $a(z)$ as a {\it metric zero}. 
The curvature scalar  in conformal coordinates is 
\be
R = -(D-1)\frac{(D-6)(a')^2+4 a a''}{4a^3}\,, \label{eq:R_conformal}
\ee
{where the prime $(^\prime)$ denotes derivative with respect to $z$.}

We assume that the scale factor $a(z)$ goes to zero at a \textit{finite} value $z_s$. Assuming that $z_s$ is approached from above, we have $ \underset{{z\to z_s^+}}{\lim}\frac{1}{a(z)}=\infty$. 
We then use the following property: 
\global\mdfdefinestyle{EqFrame}{ linecolor=white,linewidth=3pt,
backgroundcolor=EqFrame,
leftmargin=0cm,rightmargin=0cm }
\begin{mdframed}[style=EqFrame]
\be
\textrm{ If $a(z)$ vanishes at finite $z_s$,  then:}~~
\underset{{z\to z_s^+}}{\lim}\frac{a'(z)}{a(z)}=\infty\,. 
\label{eq:aprimelimit}
\ee
\end{mdframed}

This property is shown in App.\,\ref{app:aprimelimit}
using the mean value theorem --- with the mild assumption that $\frac{a'(z)}{a(z)}$ is strictly monotonic in the vicinity of the singularity. 
Using the definition of the curvature \eqref{eq:R_conformal}, \eqref{eq:aprimelimit} implies that $\underset{{z\to z_s^+}}{\lim} R(z) =\infty$. 
That is, the scalar curvature diverges at $z_s$.  
Summarizing:
\begin{mdframed}[style=EqFrame] 
\be
\shortstack[l]{
\textrm{If there is a conformal coordinate $z_s < \infty $ for which the scale factor} \\ 
\textrm{satisfies $a(z_s)=0$, then there is  a curvature singularity at $z_s$.
}
}
\hspace{-1cm}\label{eq:prop_sing}
\ee
\end{mdframed}
We emphasize that the assumption  of \textit{finite} $z_s$ is key for the proof. If $z_s=\infty$, the mean value theorem does not apply, and thus Prop.~\eqref{eq:aprimelimit} does not hold. As an example of non-applicability, AdS space in conformal coordinates $a(z)\propto z^{-2}$, {for which $z_s=\infty$,} has {$a(z_s)=0$}, yet it has finite $R$ at $z_s$ since $R$ is constant everywhere.

Finally, when a singularity appears, the general relativity description breaks down and we must cut off spacetime. In our case we have to require that $z>z_s$, which is a regular boundary. 
In other words, the metric zero describes a singularity that truncates spacetime.

\subsubsection{The  $\Mnu$ spacetime case }

We summarize the structure of boundaries and singularities of the $\Mnu$ spacetime, bearing in mind the  general discussion of Sec.~\ref{se:metric_zeros}.  

In $\Mnu$, the scalar curvature diverges at $r\to0$ for any  $\nu>0$.  
There is thus a curvature singularity at $r=0$, which  lies in the $\MnuM$ part of the spacetime. 
The singularity is labeled as ``good'' in the sense of Refs.~\cite{Gubser:2000nd,Cabrer:2009we} 
if $\nu < \sqrt{d}$.  This is confirmed by \cite{Fichet:2023dju} {for $d=4$} where it is found that  $\nu\in[0,2)$ is the range of values for which the singularity can get censored by a black hole horizon. Regarding the $\MnuP$ spacetime, it does not feature any singularity for any $\nu$. 

We now go to conformal coordinates to analyze boundaries, for which $a(z)\propto z^{\frac{2}{\nu^2-1}}$, see \cite{Fichet:2023dju} for details. 
For $\nu < 1$  there is a conformal timelike boundary, however it belongs to $\MnuP$. 
For $\nu >1$ there is a regular boundary at $z=0$. 
In the special case of $\nu=1$ {(linear dilaton)} the boundaries are null, the Penrose diagram is the same as Minkowski's, and the singularity is at a spatial infinity of the causal diamond.

We can see that Prop.\,\eqref{eq:prop_sing} applies. 
For $\nu>1$, we have $a(z_s)=0$ at a finite conformal distance $z_s=0$. 
Using \eqref{eq:prop_sing}  we conclude that there is a curvature singularity there, that coincides with the regular boundary. That is, the singularity cuts off space to $z>0$. 
For $\nu\leq 1$, we have instead $z_s=\infty$. In that case Prop.\,\eqref{eq:prop_sing} does not apply: there is a singularity for $\nu>0$ but not when $\nu=0$ {(AdS case)}.

\section{Holographic Confinement}
\label{se:string}

We classify the $\Mnu^-$ dilaton gravity backgrounds in terms of confinement. 
In the holographic view of confinement, a Wilson loop on the boundary is evaluated as a string  extending into the bulk~\footnote{See e.g.~\cite{Witten:1997ep,Hanany:1997hr,Rey:1998ik,Maldacena:1998im,Polyakov:1998ju,Kinar:1998vq, Kinar:1998pj} for seminal papers and \cite{Gursoy:2007er} for a detailed study in holographic QCD. Here we mostly  use~\cite{Kinar:1998vq,Gursoy:2007er}.  }. 
The background is said to be \textit{confining} if the  Wilson loop features an area law, i.e.~if the corresponding $q\bar q$ potential grows linearly with $q\bar q$ separation. 
In the $\Mnu^-$ dilaton gravity background  considered here, the string is attached to the brane located at $r=r_b$.

\subsection{The Holographic Wilson Loop}

\label{se:wilson_loop}

We review the holograhic computation of the quark-antiquark potential, here generalized to any bulk dimension. 
The potential energy $V(\ell)$  of a static $q\bar q$ pair separated by a distance $\ell$ is conjectured to be computed in the bulk by 
\be
T V(\ell)\equiv S_{\rm NG}[x^M_{\rm cl}(\sigma,\tau)]\,.
\label{eq:def_V}
\ee
$S_{\rm NG}$ is the Nambu-Goto action, $x^M(\sigma,\tau)$ is 
the worldsheet embedding for the string attached to a rectangle of size $(T,\ell)$ on the brane, and $x^M_{\rm cl}(\sigma,\tau)$ describes the string configuration of minimal surface, that we refer to as the classical configuration. The Nambu-Goto action in \eqref{eq:def_V} is ``on-shell'' in the sense that it is evaluated on the classical configuration. 

 We have
\be
S_{\rm NG}[x^M(\sigma,\tau)]= T_{\fund} \int d\tau d\sigma
\sqrt{{\rm det}(-\hat g_{\rm s})}
\ee
{where $\hat g_s$ denotes the induced metric on the worldsheet}, and {$T_{\fund}$} is the string tension. 
The worldsheet embedding $x^M(\sigma,\tau)$ defines the $D$-dimensional string frame metric $(g_s)_{MN}$ as
\be
(\hat g_s)_{\alpha\beta}= (g_s)_{MN} \partial_\alpha x^M 
\partial_\beta x^N\,.  \label{eq:gsgs}
\ee
Assuming that the  string frame metric takes the diagonal form \be
ds^2_{\s} = -g_{{\s},00} dt^2 +g_{{\s},rr}dr^2 + g_{{\s},ii}dx_i^2\,,
\ee
 choosing worldsheet coordinates $\tau=t$, $\sigma=x^1\equiv x$, and assuming invariance in time,  one obtains~\cite{Kinar:1998vq}
\be
 S_{\rm NG}[x^M(\sigma,\tau)] =  T_{\fund} \int dx \sqrt{h^2(r)+k^2(r)r'^2 }\,, \label{eq:SNG_simp}
\ee
with
\be
\quad  \f^2(r)=g_{{\s},00}(r)g_{{\s},ii}(r)\,,\quad \g^2(r)=g_{{\s},00}(r)g_{{\s},rr}(r) \,,
\ee
and  $r'\equiv \frac{dr}{dx}$. The classical solution, i.e.~the configuration of minimal length, is denoted by $r_{\rm cl}(x)$.

Assuming no restriction on the domain of coordinates, the string geodesic equation that determines $r_{\rm cl}(x)$ can be readily deduced from Eq.\,\eqref{eq:SNG_simp}, 
 \be
 \frac{dr}{dx}=\pm \frac{\f(r)}{\g(r)}\frac{\sqrt{\f^2(r)-\f_0^2}}{\f_0}  \,,
 \label{eq:geod_string}
 \ee 
with $\f_0=\f(r_0)$, where $r_0$ corresponds to the tip of the geodesic which satisfies $\frac{dr}{dx}\big|_{r_0}=0$. $\f_0$ can be considered as a constant of motion that parameterizes the geodesics.

 Finally, for a given spacetime background,  the string frame metric $(g_s)_{MN}$ is obtained by converting frames in the low-energy effective action. The string and Einstein frames are  related in any dimension by use of the conformal rescaling
\be
g_{MN}= e^{-2\bar \phi}g_{{\rm s},MN} \,,
\ee
where $\bar\phi = \frac{\phi}{\sqrt{(D-2)M^{D-2}_D}}$ is the reduced dilaton field introduced in section~\ref{se:setup}. 
Starting from  the Einstein frame action \eqref{eq:action}, 
we find the bulk action in string frame 
\be
\S_{\s} =  \frac{M_D^{D-2}}{2} \int d^Dx  \sqrt{g_{\rm s}}\, e^{-(D-2)\bar\phi}  \bigg(  {}^{(D)}R_{\s}  + (D-2)^2 (\partial_M \bar \phi)^2 - \bar V_{\s}(\bar\phi)  \bigg) \,,
\ee
where $\bar V_{\s} = 2 (D-2) e^{-2\bar\phi} \bar V$. The low-energy string effective action  in the usual convention from e.g.~\cite{Fradkin:1985ys,Callan:1985ia,Lovelace:1986kr} is recovered by normalizing the dilaton field as $\bar \phi_{\s}\equiv \frac{D-2}{2} \bar\phi$.

\subsection{String Shape  in Dilatonic Background}

We apply the above formalism to our dilatonic background $\Mnu$ with
Einstein  metric \eqref{eq:ds2_Mnu}. 
The string frame metric and derived quantities are
\begin{eqnarray}
ds^2_{\nu,r_b,\s} &=& e^{2\bar \phi_b} \left( \frac{r_b}{L}\right)^2 \left(\frac{r}{r_b}\right)^{2-2\nu} \eta_{\mu\nu} dx^\mu dx^\nu +
e^{2\bar \phi_b} 
 \frac{1}{\eta^2 r_b^2 } \left(\frac{r}{r_b}\right)^{2\nu^2-2\nu-2} 
dr^2  \,,  \\
\f(r) &=&  e^{2\bar \phi_b} \left( \frac{r_b}{L}\right)^2 \left(\frac{r}{r_b}\right)^{2-2\nu}\,,\quad\quad 
\g(r)=  e^{2\bar \phi_b} \frac{1}{\eta L} \left(\frac{r}{r_b} \right)^{\nu^2-2\nu}\,. \label{eq:fgstring}
\end{eqnarray}

We can notice that the metric is Minkowski when $\nu=1$: the linear dilaton background is flat in the string frame. 
The reduced potential  is  $\bar V_{\s} = -(D-2)(D-1-\nu^2)k^2 e^{2(\nu-1)\bar\phi}$. In the linear dilaton case, $\bar V_{\s}$ reduces to a constant,  which is  consistent with results in the literature. The value is $\bar V_{\s}|_{\nu=1}=-(D-2)^2 k^2$. 

To understand the behavior of the string in our dilatonic background, let us consider a static string configuration with endpoints lying at coordinates $(0,\tilde r)$ and $(\Delta x,\tilde r)$ in the bulk. 
In the critical case $\nu=1$ i.e.~the linear dilaton, the classical string configuration  simply is the straigth line, $r_{\rm cl}(x)\equiv\tilde r$. 
We establish a useful qualitative property: 
\begin{mdframed}[style=EqFrame] 
 \be
\shortstack[l]{ \textrm{If $\nu<1$ (resp. $\nu>1$) the classical string configuration bends } \\ \hspace{0cm}\textrm{towards $r<\tilde r$ (resp. $r>\tilde r$). }
       } \label{eq:string_property}
 \ee
 \end{mdframed}
Schematically,  $\vcenter{\hbox{\includegraphics[width=0.2\linewidth,trim={10cm 9.5cm 11cm 9cm},clip]{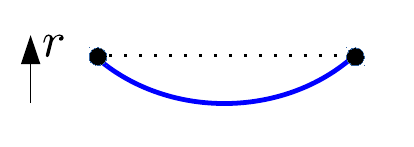} }}$\,for $\nu<1$ and $\vcenter{\hbox{\includegraphics[width=0.2\linewidth,trim={10cm 9.5cm 11cm 9cm},clip]{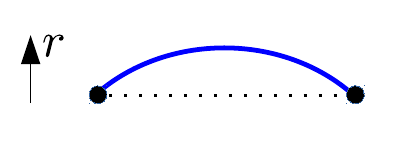} }}$ for $\nu>1$.

We show this at the infinitesimal level as follows. We consider the straigth line configuration, given by $r(x)\equiv \tilde r$ for all $x$. 
The corresponding Nambu-Goto action is $ \tilde {\cal S}_{\rm NG}=T_{\fund} \Delta x\, \f(\tilde r)$. 
We then consider a small perturbation to the straight line, $r(x)=\tilde r+\delta r(x)$, with $\delta r \ll \tilde r$ for all $x$. The action for this perturbed configuration is 
\be
  {\cal S}_{\rm NG}=  \tilde {\cal S}_{\rm NG} + \delta  {\cal S}_{\rm NG} \,,\quad \quad 
 \delta{\cal S}_{\rm NG} =T_{\fund} \int dx  \left( \f'(\tilde r)\delta r(x) +O(\delta r^2)\right) \,,
\ee
with $\f'(r)=\frac{d\f(r)}{dr}$. 
Let us further assume that the perturbation is towards $r>\tilde r$, i.e.~$\delta r(x)>0$ for all $x$. It follows that \be {\rm sign} (\delta{\cal S}_{\rm NG})= {\rm sign} (\f'(\tilde r))\,. \ee 
This relation implies that, relative to the straight line configuration, the perturbed configuration has lower (higher) energy if $\f'(\tilde r)<0$ ($\f'(\tilde r)>0$).   Similarly, for $\delta r(x)<0$ {(in which case the perturbation is toward $r<\tilde r$)} the perturbed configuration has lower energy if $\f'(\tilde r)>0$.

Using the explicit expression of $\f$  in Eq.~\eqref{eq:fgstring}, it follows that, at the infinitesimal level, string configurations  with $r<\tilde r$ (resp.~$r>\tilde r$) are favored if $\nu<1$ (resp.~$\nu>1$). This proves the infinitesimal version of Prop.\,\eqref{eq:string_property}.\,\footnote{
One may  consider the most general configuration for which $\delta r(x)>0$ on some (possibly disjoint) subdomain $I_+$ of $(0,\Delta x)$ and $\delta r(x)<0$ on its complement $I_-$. In this case the analysis can be  applied piecewise to each  subdomain where $\delta r(x)$ has  definite sign. Assuming for example {$\nu > 1$}, we know that this configuration has higher energy than a configuration that is identical on $I_+$ and has $\delta r(x)>0$ on $I_-$ i.e.~that has $\delta r(x)>0$ everywhere. This is true for any configuration, hence 
the configuration of lower energy necessarily has $\delta r(x)$ of definite sign over the whole  $(0,\Delta x)$ interval. It is thus sufficient to focus on this case. 
} 

We can also derive Prop.\,\eqref{eq:string_property} at the finite level by inspecting the classical string configurations. Assuming that the tip of the classical configuration $r_0$ is inside the spacetime,
the classical configuration is determined by the geodesic equation \eqref{eq:geod_string}. 
 We see from this equation that $\f_0$ must be smaller than $\f(r_{\rm cl})$ for all values of $r_{\rm cl}$. This condition is satisfied in the  $r_0<r_{\rm cl}< \tilde r$ region if $\f$ is monotonic  increasing, and conversely in the $r_0>r_{\rm cl}>\tilde r $  region if $\f$ is monotonic decreasing. Since $\f$ increases if $\nu<1$, and decreases if $\nu>1$, we obtain again Prop.\,\eqref{eq:string_property}.

 Finally, we notice that the scalar curvature in string frame is 
\be
R_s=-(D-1)(\nu-1)^2 (D+2\nu)  e^{-2\bar v_b } \eta^2 \left(\frac{r}{r_b}\right)^{2\nu(1-\nu)}\,.
\ee
Hence the curvature singularity in string frame is at $r\to \infty$ for $\nu<1$ and $r=0$ for $\nu>1$. Hence qualitatively we can say that, the string living in the $\Mnu$ background is \textit{repelled} by the (string frame) singularity. Accordingly, for $\nu=1$ we have $R_s=0$, hence there is no singularity, therefore the string is straight.

\subsection{Confinement}

\label{se:confinement}

\begin{figure}[t]
\centering
\includegraphics[trim={5.6cm 5cm 9.5cm 7cm},clip,width=0.37\textwidth]{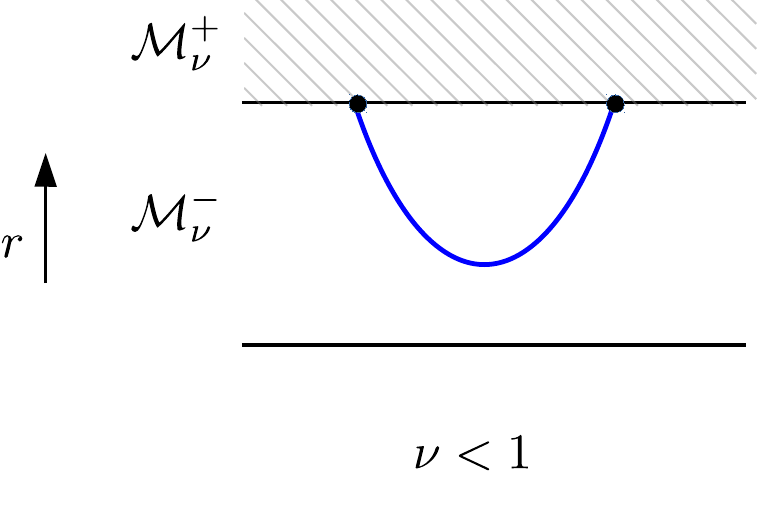}
\includegraphics[trim={8cm 5cm 9.5cm 7cm},clip,width=0.3\textwidth]{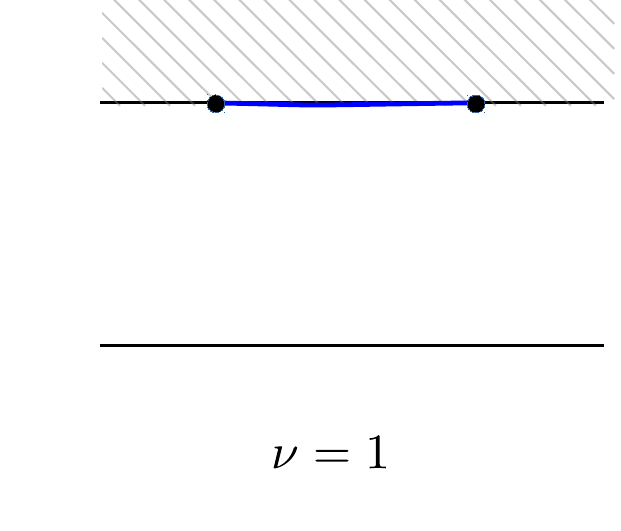}
\includegraphics[trim={8cm 5cm 9.5cm 7cm},clip,width=0.3\textwidth]{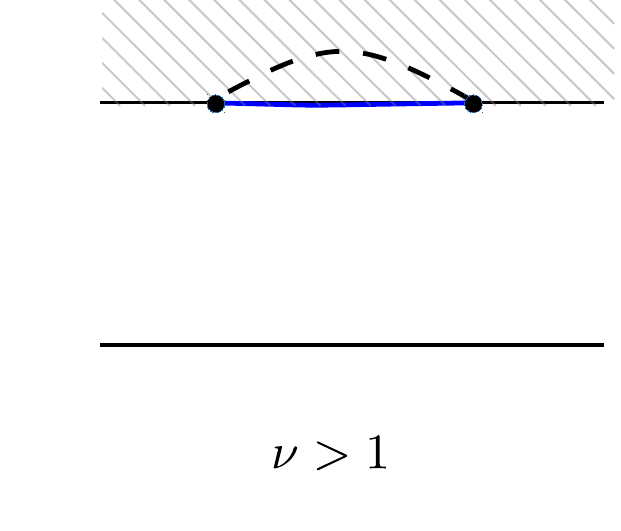}
\caption{\it   The classical string attached to the brane in the $\MnuM$ background. The configuration is a straight line  for both $\nu=1$ (because it is geodesic) and $\nu>1$ (because it is stuck to the brane). In both cases the background is confining with same string tension {$T_{\rm s} = T_{\fund} \, e^{2\bar\phi_b}\left(\frac{r_b}{L}\right)^2$. 
}}
\label{fig:string}
\end{figure}

We consider the $\MnuM$ spacetime, for which $0<r<r_b$.  The string is anchored to the brane at $r=r_b$. The behavior of the classical configuration is obtained using Prop.~\eqref{eq:string_property}.

If $\nu<1$, the classical string configuration  bends towards $r<r_b$.
In that case the geodesic equation \eqref{eq:geod_string} applies. 
The string length is given by \cite{Kinar:1998vq} 
\be
\ell = \int_{\rm cl} dx = 2\int^{r_b}_{r_0} dr\, \frac{\g(r)}{\f(r)}\frac{\f_0}{\sqrt{\f^2(r)-\f_0^2}}  \,, 
 \label{eq:l_string}
\ee
and the potential is given by 
\be
V(\ell) = T_{\fund} \int_{\rm cl} {\cal L} dx = 2  T_{\fund} \int^{r_b}_{r_0} dr\, \frac{\g(r)}{\f(r)}\frac{\f^2(r)}{\sqrt{\f^2(r)-\f_0^2}}  \,.
\label{eq:V_string}
\ee
The integrals can be done analytically, but the expressions are not particularly enlightening. 
We find numerically that $V(\ell)$ is not linearly proportional to $\ell$ for any value of $r_0$, except in the $\nu\to 1$ limit.

If $\nu=1$, the classical string follows a straight geodesic along $r=r_b$. We find
\be V(\ell)=T_{\fund}\, e^{2\bar\phi_b}\left(\frac{r_b}{L}\right)^2\ell\,. \label{eq:V_conf}
\ee 
That is, the potential grows linearly with $\ell$, for any $\ell$. The linear dilaton is therefore a confining background with
string tension 
\be
T_{\rm s}=T_{\fund}\, e^{2\bar\phi_b}\left(\frac{r_b}{L}\right)^2\,. 
\label{eq:T_conf}
\ee
The same result is obtained when taking the $\nu\to 1$ limit from \eqref{eq:l_string} and \eqref{eq:V_string}.\,\footnote{{A related analysis of confinement from the viewpoint of string theory can be found in \cite{Faedo:2017fbv}}.}

   If $\nu>1$, the string would bend towards $r>r_b$, however this region does not exist {in $\mathcal M_\nu^-$}. Using \eqref{eq:string_property}, the allowed configuration of lower energy is the straigth line along $r=r_b$. 
We find again the confining potential Eq.\,\eqref{eq:V_conf} because $\f(r_b)$ is independent on $\nu$, hence $\f(r_b)|_{\nu>1}=\f(r_b)|_{\nu=1}$. The string tension is again given by Eq.\,\eqref{eq:T_conf}. 
These features are summarized in Fig.\,\ref{fig:string}.

In terms of the curvature singularity in string frame, we can  say that confinement occurs if the singularity is in $\MnuM$, such as the string is repelled as shown in the $\nu>1$ case of  Fig.\,\ref{fig:string}, or if it is absent  such that the string remains straight as in the $\nu=1$ case of  Fig.\,\ref{fig:string}.

\subsection{Comparison to Asymptotically AdS backgrounds}

The remarkably simple picture of confinement obtained here is consistent with the one derived in~\cite{Gursoy:2007er}.
Ref.~\cite{Gursoy:2007er} considers a class of 5D dilatonic spacetimes which is asymptotically AdS in the UV and analogous to the $\Mnu$ background in the IR.
In our coordinates the AdS region is at large $r$, hence the IR region is similar to the $\MnuM$ background considered here. Roughly speaking,
spacetime continues into asymptotic AdS  instead of stopping at the brane. 
The behavior of the string in the asymptotically AdS backgrounds can be easily understood using Prop.\,\eqref{eq:string_property}. 

\begin{figure}[t]
\centering
\includegraphics[trim={5cm 5cm 9.5cm 5cm},clip,width=0.41\textwidth]{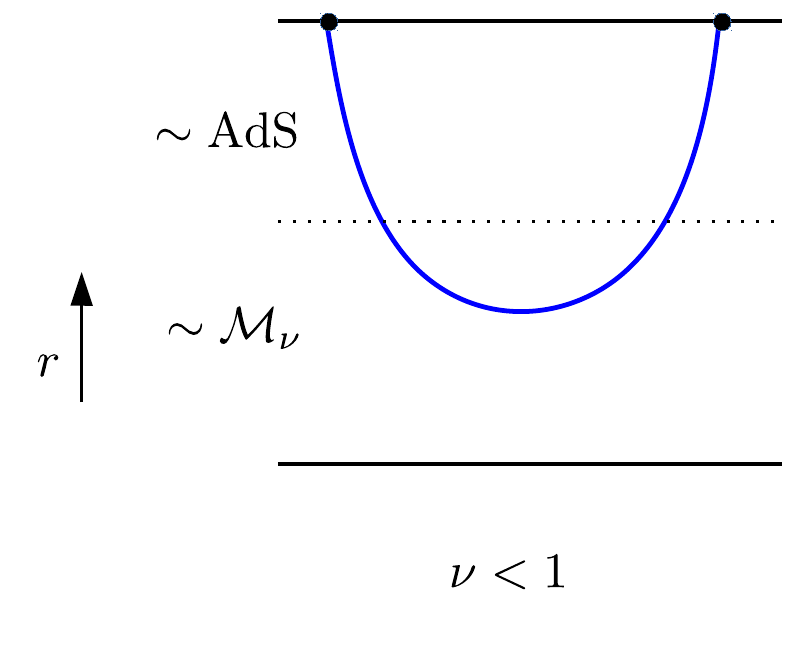} 
\hspace{1.5cm}
\includegraphics[trim={7cm 5cm 9.5cm 5cm},clip,width=0.35\textwidth]{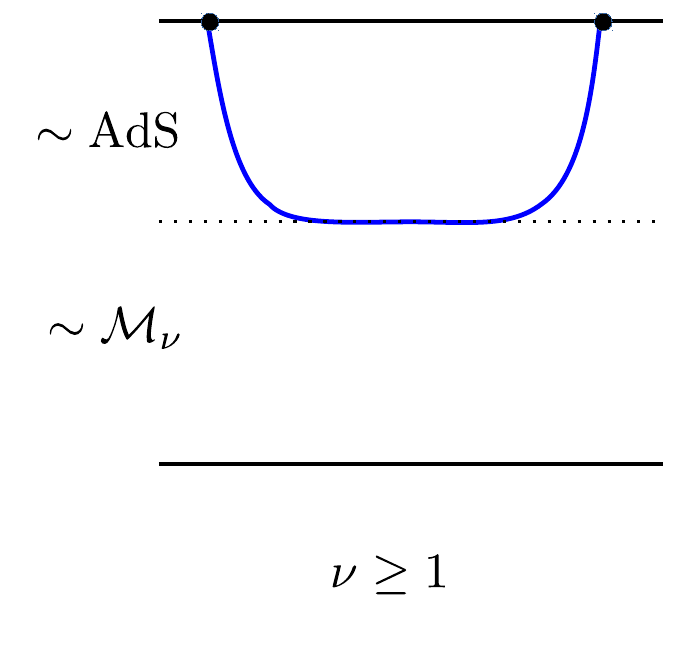}
\caption{\it   The classical string attached to a brane in an asymptotic AdS background. 
For $\nu\geq 1$ the string enters the asymptotic $\Mnu$ region only up to a finite value of $r$. Accordingly it asymptotically 
follows an area law with tension $T_{\rm s} = T_{\s} \, e^{2\bar\phi_b}\left(\frac{r_b}{L}\right)^2$. 
}
\label{fig:stringAdS}
\end{figure}

Let us consider a string anchored to a brane in the AdS region, i.e.~at large $r$, with endpoints separation $\Delta x$ in Minkowski distance.\,\footnote{{Since the asymptotic AdS spaces considered here are cutoff by the brane,  the string self-energy \cite{Schwarz:1982jn,Green:1987mn,Chialva:2009pg} is regularized \textcolor{blue} it does not feature UV divergences. }}
For small $\Delta x$, the string only knows about the AdS background
and thus  bends towards smaller $r$, as dictated by Prop.\,\eqref{eq:string_property} with $\nu=0$.
Increasing $\Delta x$,  the tip of the string enters  in the non-AdS IR region. In this regime, if $\nu<1$, the string keeps entering further into the IR  since it still tends to bend towards smaller $r$. But if, instead, $\nu>1$,  the part of the string in the IR wants to bend towards large $r$ as dictated by \eqref{eq:string_property}.
That is, for $\nu>1$ the tendencies in the UV and IR regions are opposite. As a result the string does not go beyond a certain point in the IR. These behaviors are sketched in Fig.\,\ref{fig:stringAdS}.

This phenomenon of saturation  matches the behavior reported in \cite{Gursoy:2007er}, which was identified in terms of a non-monotonicity in the scale factor. 
It turns out that this saturation  implies confinement asymptotically. 
The confining behavior is easily understood 
from the viewpoint of our analysis, 
for example by
replacing  the shaded region in Fig.\,\ref{fig:string} by asymptotic AdS.
At large $\Delta x$,
for $\nu\geq 1$ the main contribution to the string length would be the straight segment, that implies confinement as shown in section \ref{se:confinement}.

\section{Holographic Entanglement Entropy  }
\label{se:EE}

In a given quantum system, the entanglement of a given subsystem with the rest of the system can be quantified using the entropy. This entanglement (or von-Neumann) entropy {(EE)} is defined  via tracing out the states outside the subsystem. It can  be viewed as the amount of entropy for an observer that  only receives information from the  subsystem. Entanglement entropy  is typically used as a tool to understand quantum systems.

The entanglement entropy in holographic theories  admits a simple geometric description~\cite{Ryu:2006bv,Nishioka:2009un,Casini:2011kv,Headrick:2013zda,Blanco:2013joa,Hubeny:2013gta,Faulkner:2013ana,Lewkowycz:2013nqa,Barrella:2013wja,Engelhardt:2014gca}. 
The  interplay of holographic entanglement entropy (HEE) with confinement has been explored in~\cite{Klebanov:2007ws,Bah:2007kcs,Fujita:2008zv,Engelhardt:2013jda,
Chakraborty:2018kpr,Fujita:2020qvp,daRocha:2021xwq,Jokela:2023lvr,
Kol:2014nqa,Jokela:2020wgs,Grieninger:2023pyb,Fatemiabhari:2024aua}.  Here our goal is to understand HEE in the simple setup of dilaton gravity. 
We explore the HEE described by the $\MnuM$ background, and also establish some properties of HEE for more general warped metrics.

The subsystem of our interest is supported on a region $A$ of the brane. Focusing  on a spacelike region,  the HEE is given by the Ryu-Takayanagy (RT) formula~\cite{Ryu:2006bv}
\be
    S_A = \frac{\mathrm{Area}(\gamma_A)}{4 G_N} \,,
    \ee   
    where $\gamma_A$ is the  surface of minimal area anchored to the boundary of $A$.~\footnote{ 
    The RT formula should generally be accompanied by a homology condition that dictates how the minimal surface behaves with respect to black holes (see~\cite{Asplund:2014coa,Almheiri:2016blp,Neuenfeld:2021bsb}). 
    In this section we do not consider black holes, hence the bulk has trivial homology and no additional conditions need to be specified.}  {The area  of $\gamma_A$ is computed using the Einstein frame metric.}
    
    {In this work}
     we choose the simplest region for $A$: a   $(d-1)$-dimensional strip with width $\L_A$ at a given time $t=0$.

In the following we first establish some properties
for the general warped metric defined in \eqref{eq:ds2_gen}, before turning to the detailed analysis of the $\MnuM$ background.

\subsection{Elementary Properties }

\begin{figure}[t]
\centering
\includegraphics[trim={2.5cm 6cm 6.5cm 2cm},clip,width=0.46\textwidth]{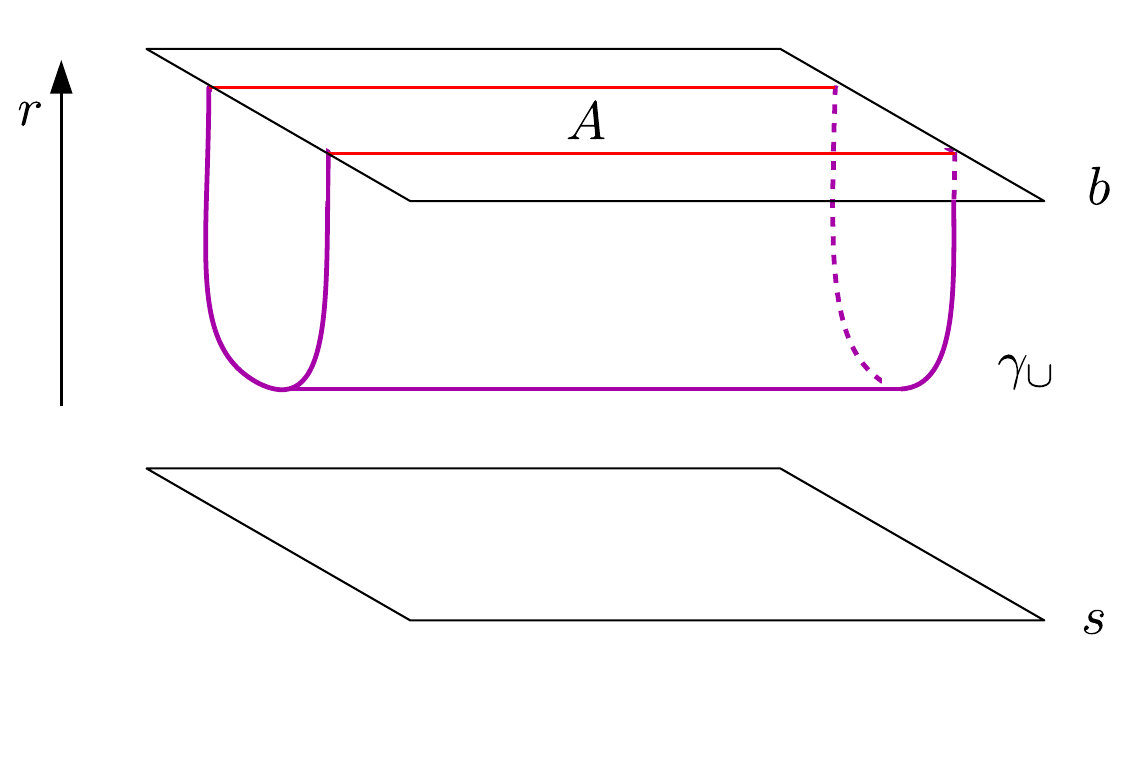} 
\hspace{0.7cm}
\includegraphics[trim={2.5cm 6cm 6.5cm 2cm},clip,width=0.46\textwidth]{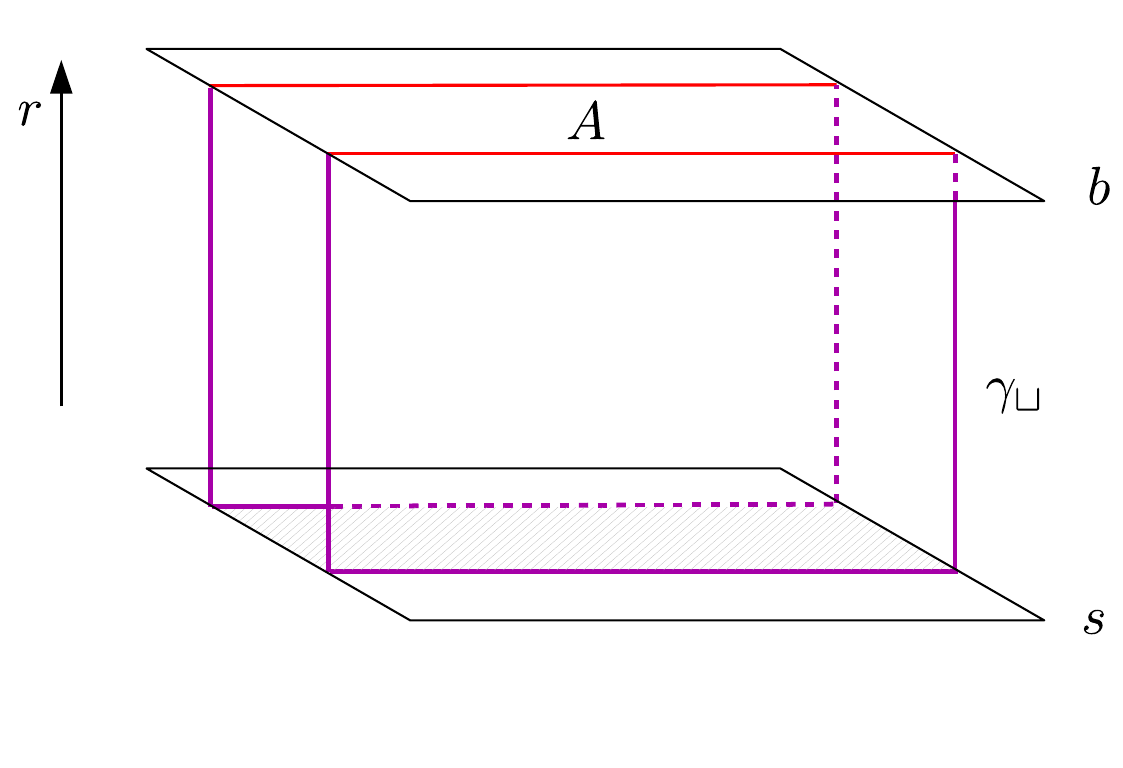}
\caption{\it   
The candidate minimal surfaces anchored to the surface of a $(d-1)$-strip  at a given time on the brane. Left: The smooth surface. Right: The square surface.  
}
\label{fig:surfaces}
\end{figure}

We can identify two possible surfaces: \textit{i)} a smooth surface generated by the shift of a curve with base $\Delta x = \L_A$, and \textit{ii)} a squared-shape surface that extends from the brane ($r=r_b$) down to  $r=0$. The surfaces are respectively denoted $\gamma_{\cup}$ and $\gamma_{\sqcup}$ and are depicted in Fig.\,\ref{fig:surfaces}. 

We are interested in the behavior of $S_A$ as a function of the strip width $\L_A$. 
Anticipating the  HEE results, the existence of these two surfaces is the cause of \textit{phase transitions} in the EE, which happen because either $\gamma_{\cup}$ or  $\gamma_{\sqcup}$ have minimal area depending on the subsystem size.

The surfaces in Fig.\,\ref{fig:surfaces} are  parametrized by an embedding $x^M=F(\sigma^m)$ where $\sigma^m$ are the $d$ coordinates for the points belonging to the surface. We choose  $F(\sigma^m)=(\sigma^m,u(\sigma^1) )$, such that the $\sigma^m$ are identified with $(r,  x^1, \cdots, x^{d-2})$. 
  In Fig.~\ref{fig:surfaces}, $x^{d-1}$  is the direction  which crosses the strip. The profile of the minimal surface is given by $x^{d-1}=u(r)$. The vertical sides of the square surface correspond to  $x^{d-1} = \mathrm{cst}$. 
 The longitudinal directions collectively represent the other $(x^1,\ldots,x^{d-2})$ directions. 
 {Working in the Einstein frame with the general metric~\eqref{eq:ds2_gen}},
 the induced metric for the surface is then
\be
    ds_{\rm ind}^2 = \Big(g_{r r}(r) + u^{\prime\, 2}(r)g_{x x}(r) \Big)dr^2 + g_{x x}(r)\Big( (dx^1)^2 + \cdots + (dx^{d-2})^2 \Big) \,. \label{eq: induced_metric}
\ee

\subsubsection{Smooth Surface}

From the induced metric \eqref{eq: induced_metric}, we can compute the spatial area of the smooth surface as
\be
    \mathrm{Area}(\gamma_{\cup}) = 2 A_{d-2}\int_{r_0}^{r_b} dr \, a(r)^{\frac{d-1}{2}} \sqrt{b^2(r) + u^{\prime\, 2}(r)}\,, \label{eq: area_smooth} 
\ee
with  $ a(r)= g_{xx}(r)$, $ b(r)=\sqrt{\frac{g_{rr}(r)}{g_{xx}(r)}}$ and  $A_{d-2} = \int dx^1  \cdots dx^{d-2}$. 

The surface is minimal when the   $u(r)$ function  satisfies the Euler-Lagrange equation 
\be
    \frac{d u(r)}{dr} = U \frac{b(r)}{\sqrt{a(r)^{d-1}-U^2}}\,.
    \label{eq: f_equation} 
\ee
Here $U$ is a constant that characterizes the curve $x^{d-1} = u(r)$.  Since $\frac{dr}{d u}$ vanishes if $a^{d-1}(r)= U^2$, the $U$ constant is identified as $U \equiv a^{\frac{d-1}{2}}(r_0) $ where $r_0$ is the coordinate of the ``tip'' of the curve. 
The strip width $\L_A$ and the tip of the curve $r_0$ are tied due to the solutions of \eqref{eq: f_equation}, i.e.~we have $\L_A=\L_A(r_0)$.

\subsubsection{Square Surface and Singularity}\label{se:existence_strip}

The area of the square surface is the one of the vertical piece plus the one of the strip at $r=0$. The existence of $\gamma_{\sqcup}$ requires that the 
vertical piece exist in a first place. In terms of the general warped metric \eqref{eq:ds2_gen}, this is true if there is a value of $r_0$ such that the metric coefficient $a(r)$ satisfies $a(r_0)\equiv 0$. The  bottom part of the square surface is localized at~$r_0$, which implies that its area is vanishing due to the vanishing of the scale factor. Hence we have  always that
\be
    \mathrm{Area}(\gamma_{\sqcup}) = 2 A_{d-2}\int_{0}^{r_b} dr \, a(r)^{\frac{d-1}{2}} b(r) \,,  
\ee
for \textit{any} $\L_A$,  i.e. the area of the square surface does not depend on the strip width.

 Assuming conformal coordinates $(x^\mu,z)$
 in the general metric such that $b(z)=1$ for all $z$, the existence of $\gamma_{\sqcup}$ requires $a(z_0)\equiv 0$, which implies that $z_0$ is a metric zero. Applying Prop.\,\eqref{eq:prop_sing}, this implies the existence of a curvature singularity at $z=0$, on which the square surface ends.

\subsubsection{{A Necessary Condition for HEE phase transition}}

\label{se:existence_PT}

The area for the square surface may be finite or infinite, depending on the behavior of the integral $\int_{0}^{r_b} dr a(r)^{\frac{d-1}{2}} b(r)$. If the integral diverges, the $\gamma_{\sqcup}$ has infinite area and thus  the minimal area is the one of $\gamma_{\cup}$. In that case, no phase transition can happen in the HEE. 
Therefore,
\begin{mdframed}[style=EqFrame]
\be
\shortstack[l]{
\textrm{A HEE phase transition can occur only if} \\  \textrm{the integral $\int_{0}^{r_b} dr a(r)^{\frac{d-1}{2}} b(r)$ is finite. } }
\ee
\end{mdframed}

This is a necessary condition.  
This condition is automatically satisfied if there is a metric zero at finite conformal distance. 
Thus the possibility of a
HEE phase transition 
is related via Prop.\,\eqref{eq:prop_sing} 
to the existence of a singularity at finite conformal distance.

\subsubsection{The Near-Brane Surface}

When $r_0$ is close to $r_b$,  the smooth surface stays near the brane. In that limit the equations \eqref{eq: area_smooth} and \eqref{eq: f_equation}  are dominated by the square root term $(a(r)^{d-1}-a(r_0)^{d-1})^{-1/2}$, while the other factors can be taken at $r\sim r_b$. This implies that 
\be
\mathrm{Area}(\gamma_{\cup}) \overset{r_0\sim r_b}{\approx}  A_{d-2} \L_A a(r_b)^{\frac{d-1}{2}}\,.
\label{eq:short_geodesic}
\ee

That is, the area of $\gamma_{\cup}$ tends to the area of the basis region, $\mathrm{Area}(\gamma_{\cup})\approx \mathrm{Area}(A)$ . The \eqref{eq:short_geodesic} limit can also be understood  as the area for ``small''  $\gamma_\cup$, since the approximation is valid in the small $\L_A$ limit.

\subsection{Phase Transition and   Boundaries }
\label{se:PT_boundary}

This section establishes some properties of the HEE for the general warped metric of \eqref{eq:ds2_gen}. 
All the properties obtained here appear explicitly in the case of the $\MnuM$ background presented in section \ref{se:HEEMnuM}, see in particular Fig.\,\ref{fig:HEE_MnuM}.

A necessary condition for a phase transition of HEE to occur is that the width of the base of the $\gamma_{\cup}$ surface be bounded from above,  i.e.~$\L_A<\L_A^{\rm max}$ with $\L_A^{\rm max}<\infty$. 
In such a situation, whenever one considers a strip $A$ with $\L_A>\L_A^{\rm max}$, only the $\gamma_{\sqcup}$ remains available, hence $\gamma_{\sqcup}$ is automatically the minimal surface. 
Here we establish a relation between the boundedness of the base of $\gamma_{\cup}$, i.e.~the existence of a finite $\L_A^{\rm max}$, and the existence of a singularity at finite conformal distance.

As seen in section~\ref{se:singularities}, the spacetime background may feature either a  regular or a conformal timelike boundary along constant $r$ slices. 
Since this distinction is done using conformal coordinates, here we go to conformal coordinates $(x^\mu,z)$, that imply $b(z)=1$ for any $z$ such that the warped metric \eqref{eq:ds2_gen} is manifestly conformally flat.
We assume in this subsection that the metric coefficients are \textit{strictly monotonic}.\,\footnote{
The results of this subsection can be extended with little effort to metrics that are monotonic in a finite region near the singularity. 
}

Using the Euler-Lagrange equation \eqref{eq: f_equation}, we know that the width of the base of $\gamma_{\cup}$ satisfies
\be
\L_A(z_0)= U \int_{z_0}^{z_b}  \frac{dz}{\sqrt{c(z)-U^2}} \,, \label{eq:Deltax_conformal}
\ee
where $c(z) = a(z)^{d-1}$ and $U^2=c(z_0)$, with $z_0$ the coordinate of the tip of the curve. If  $c(z)$ is monotonically increasing (resp. decreasing) we have $z_0<z_b$, (resp. $z_0>z_b$), such that $c(z)>c(z_0)$ under the integral in all cases.  In the following we choose $c$ to be increasing and thus $z_0<z_b$.

We denote the domain of $a(z)$ by $[z_*,z_b]$. 
The lower bound $z_*$ may be either finite or infinite.
 We assume in this subsection that $U$ can reach $0$, which is required for the square surface to exist (see  section~\ref{se:existence_strip}). Since $U^2=c(z)$ with $c$ monotonic, this condition corresponds necessarily to having $c(z_*)=0$.
 Using Prop.\,\eqref{eq:prop_sing}, if $z_*$ is finite then there is a curvature singularity at $z_*\equiv z_s$. If $z_*$ is infinite, no statement can be made and $z_*$ can be simply thought of as the boundary for $z$.

\subsubsection{Sufficient Conditions for HEE Phase Transition}
\label{se:condition1}

We rewrite the integral in \eqref{eq:Deltax_conformal} as
\be
\int_{U^2}^{y_b}  \frac{dy}{c'[c^{-1}(y)]\sqrt{y-U^2}} \,, \label{eq:Deltax_1}
\ee
where we used the change of variable $c(z)\equiv y$, with $y_b\equiv c(z_b)$, $y_0\equiv c(z_0)=U^2$. The function $c^{-1}$ is the inverse of the $c$ function {and $c'[x]$ is the derivative of $c$ with respect to its argument $x$}. 

Since by assumption the $c$ function is monotonic on the open interval $(z_*,z_b)$, there exists $B>0$ such that   $c'(z)>B$ for all $z\in (z_*,z_b)$. Hence 
the inverse  $1/c'(z)$ is defined on $(z_*,z_b)$. 
However it is not guaranteed that $1/c'(z)$ is defined at the endpoints, where it might blow up, e.g.~$1/c'(z_*)=\infty$. In such a case the integral \eqref{eq:Deltax_1} is improper.

A {sufficient} condition for the integral \eqref{eq:Deltax_1} to be finite is if the lower bound $z_*$ is finite and $1/c'(z)$ is defined on the \textit{closed} interval $[z_*,z_b]$. In that case, having a continuous function on a closed interval, the boundedness theorem implies that there exists an upper bound $\bar B$ on $1/|c'(z)|$ on $[z_*,z_b ]$. As a result the integral \eqref{eq:Deltax_1} is bounded from above by 
$
\bar B\int_{U^2}^{y_b}   dy (y-U^2)^{-1/2}  = 2 \bar  B  \sqrt{y_b-U^2}$.  The comparison theorem for integrals then implies that \eqref{eq:Deltax_1} is bounded from above for any $U$. 
Going back to the definition \eqref{eq:Deltax_conformal}
we use that $U$ is bounded from above since $c(z_0)<c(z_b)$ for all $z_0<z_b$ because $c$ is monotonic increasing.  It follows that  $\L_A(z_0)$ is bounded from above for any $z_0$, i.e.~there exists $\L_A^{\rm max}<\infty$.   The same  conclusion happens if one chooses the convention that $c$ is decreasing and $z_0>z_b$. Putting the pieces together, we conclude that
\be {\shortstack[l]{
\textrm{If the function $1/c'(z)$ is defined on the \textit{closed} interval $[z_*,z_b]$ with finite $z_*$, } \\ \textrm{ then the holographic entanglement entropy exhibits a phase transition. }}} \label{eq:HEEcond1}
\ee
Since having a metric zero at finite $z_*$ implies a curvature singularity by Prop.~\eqref{eq:prop_sing}, the condition in \eqref{eq:HEEcond1} requires $1/c'(z)$ to be defined on the singularity $z_*\equiv z_s$. 

We emphasize that \eqref{eq:HEEcond1} is only a sufficient condition for the existence of HEE phase transition. For example, when applied to the $\MnuM$ metric, for which $c(z)\propto z^{\frac{2(d-1)}{\nu^2-1}}$ and a regular boundary $z_s=0$ exists for $\nu>1$, we have   $1/c'(z)\propto z^{\frac{2d-1-\nu^2}{1-\nu^2}} $. This is finite at $z\to z_s$ only for $\nu>\sqrt{2d-1 }$. This condition does not ensure  the existence of the HEE  phase transition in the $\MnuM$ background for any $d$, since the absence of bad singularity requires  that  $\nu<\sqrt{d}$  (see section~\ref{se:singularities}).

{We  find a slightly better sufficient condition 
by bounding {$dc^{-1}(y)/dy\equiv[c^{-1}]'(y)= 1/c'[c^{-1}(y)]$} with the $U$-dependent upper bound $[c^{-1}]'(U^2)$. Such a bound is  set up to include the situation where $[c^{-1}]'(y)$ blows up at $y\to 0$, that is not taken into account by the sufficient condition \eqref{eq:HEEcond1}. We put this upper bound in the definition \eqref{eq:Deltax_conformal} and  require  finiteness for any $U$. We obtain that $\L_A$ is bounded from above for any $U$ (i.e.~any $z_0$) if the condition 
\be
U \times [c^{-1}]'(U^2)<\infty \, \label{eq:cond2a}
\ee
is satisfied for all $U$.  Using  $U^2=c(z_0)$ and $c(z)=a(z)^{d-1}$, we obtain the final condition that is summarized as 
\begin{mdframed}[style=EqFrame]
 \be
 \shortstack[l]{
\textrm{
{If} the function $\frac{a^{\frac{3-d}{2}}(z_0)}{a'(z_0)}$ is bounded from above for all $z_0\in (z_*,z_b]$, } \\ \textrm{ then the  HEE exhibits a phase transition.  } 
} 
\label{eq:HEEcond2}
 \ee
 \end{mdframed}
 When applied to the $\MnuM$ metric, the criterion from  \eqref{eq:HEEcond2}, is  
$\nu>\sqrt{d}$. While it is better than property \eqref{eq:HEEcond1}, the value lies right at the consistency limit  for bad singularities, hence the sufficient condition \eqref{eq:HEEcond2} 
does not ensure  the existence of the HEE  phase transition in the $\MnuM$ background for any $d$.

\subsubsection{Pairs of Smooth Surfaces}

In the presence of the  singularity at finite conformal distance, the fact that $\L_A(z_0)$ tends to zero when $z_0\to z_s$ implies that there exist \textit{two} smooth surfaces $\gamma_{\cup}$,
that both solve the Euler-Lagrange equation, for a given $\L_A<\L_A^{\rm max}$. 
This is because, for a given $\L_A$, in addition to the large surface with $z_0\sim z_s$, there is always a small surface closer to the brane, whose area is given by \eqref{eq:short_geodesic}. 
By continuity of $\L_A$ in $z_0$, for any $\L_A<\L_A^{\rm max}$ there must exist two smooth surfaces with different $z_0$.  

In summary, the presence of a singularity at finite conformal distance implies  the existence of a pair of smooth surfaces.

\subsection{Smooth Surfaces in the $\MnuM$ background}

\label{se:surfaces_Mnu}

In the $\MnuM$ background we can actually compute the smooth surfaces.
{Using the Einstein frame metric \eqref{eq:ds2_Mnu}, the $a$, $b$ functions are \be a(r)=\frac{r^2}{L^2}\,,\quad b(r) = \frac{r^{\nu^2-2}}{r_b^{\nu^2} } \frac{L}{\eta} \,.  \ee}
The general solution to the Euler-Lagrange equation \eqref{eq: f_equation} is
\begin{equation}
    u = \frac{i L^2 U }{\eta (1-\bar\nu^2) (d-1)} 
\frac{r^{\nu^2-1}}{r_0^{d-1} r_b^{\nu^2}}
    {}_2 F_1 \left(\frac{1}{2}, \frac{1}{2}(\bar\nu^2 - 1); \frac{1}{2}(\bar\nu^2 + 1); \frac{r^{2(d-1)}}{r_0^{2(d-1)}} \right) + \mathrm{cst}\,,
\label{eq:x_Mnu}
\end{equation}
where
\be
    \bar\nu^{2} = 1+\frac{\nu^2-1}{d-1} \,,
\label{eq:nuprime}
\ee
for $\nu \neq 1$ and $r_0\neq 0$. We show in App. \ref{app:app_surface} how to compute the general case with arbitrary $d$ from the $d=2$ case using the matching given in \eqref{eq:nuprime}.

The $\nu=1$ (LD)  case gives
\begin{eqnarray} u &=& \frac{U L^2}{\eta r_b r_0^{d-1} (d-1)} \arctan\left(\sqrt{\frac{r^{2(d-1)}}{r_0^{2(d-1)}}-1}\right) +{\rm cst}  
\label{eq:x_LD_sol}
\end{eqnarray}
for any $d$.

\subsubsection*{Tip}

The approximation for $r_0$ in the $r_0 \ll r_b$ limit is
\be
\frac{r_b}{r_0}\sim \left( \frac{ \eta L_A r_b}{L} \right)^{\frac{1}{1-\nu^2}} \,. \label{eq:r0_lim}
\ee
The combination $ \frac{\eta L_A r_b}{L}$ will naturally 
appear in the asymptotic and critical behaviors obtained throughout the rest of the section.

\subsubsection*{Shape}

\label{se:geodesic_shape}

The smooth surfaces have qualitatively different behaviors in the  $ \nu<1$, $\nu=1$ and $ \nu>1$ cases. 
\begin{itemize}
\item 
For $\nu<1$, the strip width $L_A$ increases indefinitely when $r_0$ approaches $0$, i.e.~we have $L_A\in[0,\infty)$.  
\item 
In contrast, when $\nu=1$  the  width reaches a saturation, i.e.~$L_A$ is bounded from above with $L_A\in[0,\frac{\pi L}{(d-1)\eta r_b}]$ (see Eq.~\eqref{eq:Area_LD} below). 
\item
Finally, for $\nu>1$, starting from $r_0\sim r_b$, the base of the smooth surface increases, reaches a maximum and then decreases to zero when $r_0$ approaches $0$. 
Thus for $\nu>1$ there exist \textit{two} smooth surfaces with same $L_A$. 
\end{itemize}

For  $\nu\leq 1$, the  smooth surfaces in the $r_0 \to 0 $ limit tend to take a square shape with a nearly flat part that approaches the $r=0$ line. They approach thus the square surface $\gamma_\sqcup$ in this limit.

\subsubsection*{Area}

\label{se:geodesic_length}

\begin{figure}[t]
    \centering
    \begin{annotatedFigure}
    {\includegraphics[trim={1.3cm -1.5cm 0 0}, clip, width=0.32\textwidth]{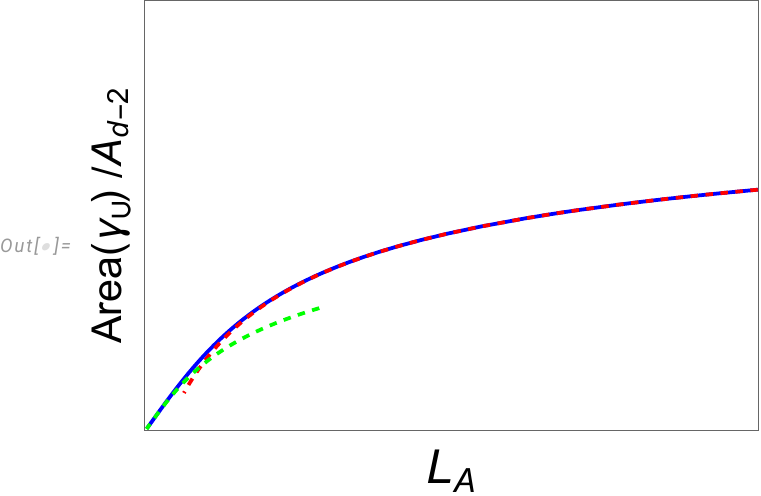}
    \includegraphics[trim={1.3cm -1.5cm 0 0}, clip, width=0.32\textwidth]{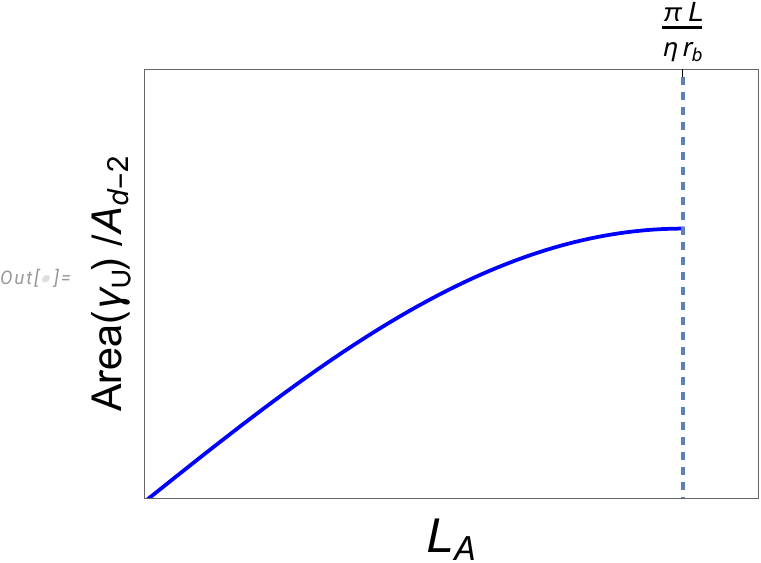}
    \includegraphics[trim={1.3cm -1.5cm 0 0}, clip, width=0.32\textwidth]{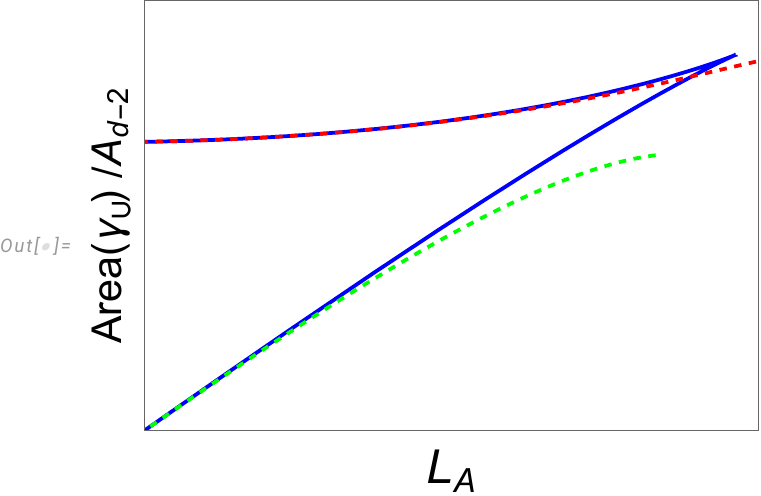}}    \annotatedFigureBox{10,0.9}{100,100}{$\nu < 1$}{0.18,0.001}    \annotatedFigureBox{0.222,0.284}{0.3743,0.4934}{$\nu = 1$}{0.52,0.001}    \annotatedFigureBox{0.222,0.284}{0.3743,0.4934}{$\nu > 1$}{0.86,0.001}
    \end{annotatedFigure}
    \caption{\it   
    Area of the smooth surface $\gamma_\cup$ as a function of the strip width $L_A$. 
Plain lines are exact results for $\nu=0.5$, $1$ and $1.5$.  Red and green dashed lines are respectively approximations for $r_0\ll r_b$ \eqref{eq:Deltas_r0small}  and for $r_0\approx r_b$  \eqref{eq:short_geodesic}. 
    }
    \label{fig:s_vs_x}
\end{figure}

The dependence of $\mathrm{Area}(\gamma_{\cup}) $ as a function of the strip width $L_A$ is shown in Fig.\,\ref{fig:s_vs_x}. 
When the surface stays near the brane, we have $r_0\sim r_b$ and the linear relation \eqref{eq:short_geodesic} appears. 
This holds true both when  $\nu\leq 1$ and for the smaller smooth surface when $\nu>1$. 

Let us study the $r_b\gg r_0$ limit. 
For  $\nu=0$ and $d=2$ (i.e.~AdS$_3$), $\mathrm{Area}(\gamma_{\cup}) $ grows logarithmically when $r_b\gg r_0$.~\footnote{{For AdS$_3$,  $\mathrm{Area}(\gamma_{\cup}) \approx \frac{2}{\eta}\log{\left(\frac{\eta L_A r_b}{L} \right)}$ in the $r_b \gg r_0$  limit. 
Using that $\eta|_{\nu=0}$  is the inverse AdS radius, and
going to Poincaré coordinates using $z=\frac{L}{\eta r}$, the familiar result from e.g.~\cite{Ryu:2006bv} is recovered, with   the AdS boundary cut off at $z_b$. }} In contrast, in all other cases, i.e.~for any
 $\nu\neq 0$ or $d\geq2$,  the area reaches a finite value given by  $\mathrm{Area}(\gamma_{\cup})|_{ r_0\to 0}= \frac{2 A_{d-2}}{  (d-2+\nu^2)\eta}\frac{r^{d-2}_b}{L^{d-2}}$. At first order we obtain the asymptotic form
valid for $r_0 \ll r_b $, 
\be 
\frac{ \mathrm{Area}(\gamma_{\cup}) }{A_{d-2}}{\frac{(d-1)L^{d-2}}{r_b^{d-2}}}\simeq 
\begin{cases} \displaystyle
   \frac{2}{\eta {\bar\nu}^2} + \frac{\sqrt{\pi} \Gamma{\left(-\frac{{\bar\nu}^2}{2} \right)}}{\eta \Gamma{\left(\frac{1}{2}(1 - {\bar\nu}^2) \right)}} \left[\frac{\eta (d-1) \Gamma{\left(\frac{1}{2}(3 - {\bar\nu}^2) \right)}}{\sqrt{\pi}\Gamma{\left(1 - \frac{{\bar\nu}^2}{2} \right)}} \frac{r_b}{L} L_A \right]^{\frac{{\bar\nu}^2}{{\bar\nu}^2 - 1}}  {\rm if} \quad {\bar\nu}^2 < 2
   \\ \displaystyle
    \label{eq:Deltas_r0small}
\frac{2}{\eta {\bar\nu}^2} + \frac{\eta (d-1)^2 ({\bar\nu}^2 -2)  r_b^2}{4 L^2}L_A^2  
 \hspace{4.5cm}{\rm if} \quad {\bar\nu}^2 > 2
 \end{cases}
\ee 
 These expressions are obtained by integrating Eqs.~\eqref{eq:area_equations_1} from $ r_0$ to $r_b$,  then solving for $r_0$ as a function of $L_A$ using the $r_b\gg r_0$ approximation. 
 In the $\nu>1$ case, these expressions describe the area of the large surfaces. 

 From \eqref{eq:r0_lim} we can see that  $r_0\ll r_b$  corresponds to $ \frac{\eta r_b L_A}{L} \gg 1$  if  $\nu<1$, and to $ \frac{\eta r_b L_A}{L} \ll  1$ if $\nu>1$. In the limits \eqref{eq:Deltas_r0small}, this implies that  the second term is always subleading and vanishes when $r_0\to0$.  For example, keeping fixed $\frac{\eta r_b}{L}$, $r_0\to0$ corresponds to either $ L_A\to\infty$ for $\nu<1$ or to $ L_A\to 0$ for $\nu>1$.  The approximations of the areas for $r_0\sim r_b$ and $r_0\ll r_b $ are shown in Fig.\,\ref{fig:s_vs_x}.

In the $\nu=1$ case, it turns out that it is possible to express the area explicitly as a function of $L_A$. We find the remarkably simple relation
\begin{equation}
   \mathrm{Area}(\gamma_{\cup})  = A_{d-2} \frac{2}{\eta (d-1)}\left(\frac{r_b}{L} \right)^{d-2} \sin{\left(\frac{(d-1)\eta r_b}{2 L}L_A \right)} \,.
   \label{eq:Area_LD}
\end{equation}
The linear dilaton  background has other geometric properties that will be further used in a future 
work. 
Here we notice that, as it can already be seen from \eqref{eq:x_LD_sol}, $L_A$ is {\it bounded from above}. 
The largest allowed $L_A$, corresponding to the argument being equal to $\pi/2$ in \eqref{eq:Area_LD}, is $L_A = \frac{\pi L}{(d-1)\eta r_b} $.

\subsection{Holographic Entanglement Entropy in the $\MnuM$ background}

\begin{figure}[h]
    \centering    \begin{annotatedFigure}
    {\includegraphics[trim={1.3cm -1.5cm 0 0}, clip, width=0.32\textwidth]{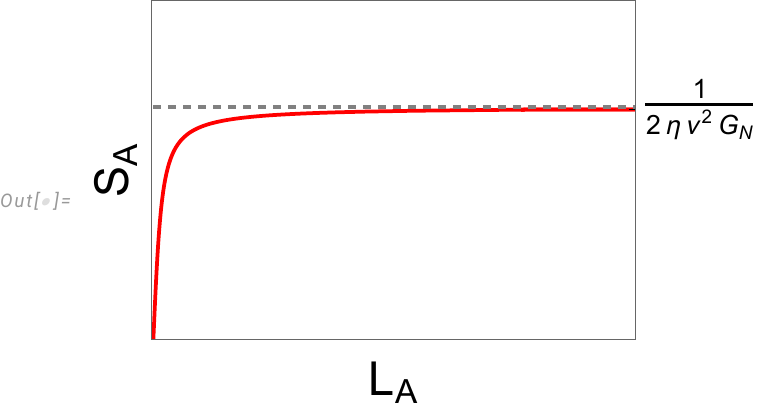}
    \includegraphics[trim={1.3cm -1.5cm 0 0}, clip, width=0.32\textwidth]{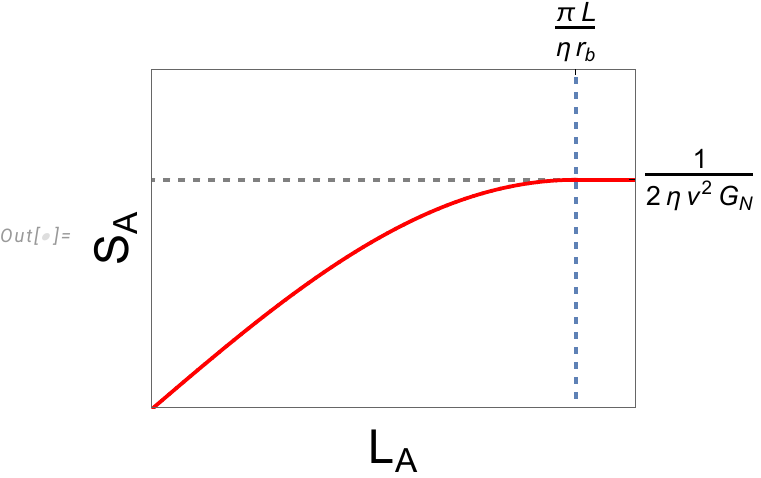}
    \includegraphics[trim={1.3cm -1.5cm 0 0}, clip, width=0.32\textwidth]{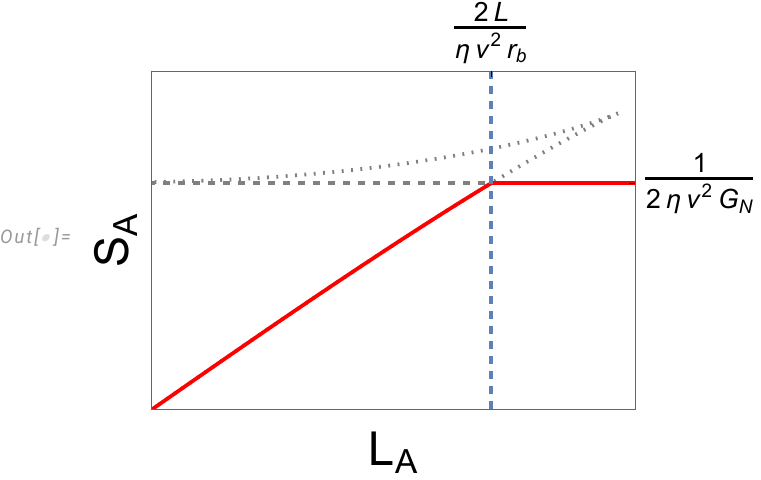}}    \annotatedFigureBox{10,0.9}{100,100}{$\nu < 1$}{0.16,0.001}    \annotatedFigureBox{0.222,0.284}{0.3743,0.4934}{$\nu = 1$}{0.49,0.001}    \annotatedFigureBox{0.222,0.284}{0.3743,0.4934}{$\nu > 1$}{0.83,0.001}
    \end{annotatedFigure}
    \caption{\it  
Holographic entanglement entropy from   the $\MnuM$ background with $d=2$ spatial dimensions, shown  as a function of the strip width. 
A second order phase transition occurs if $\nu=1$, while a first order phase transition occurs if $\nu>1$.
    }
    \label{fig:HEE_MnuM}
\end{figure}

\label{se:HEEMnuM}

We compute the holographic entanglement entropy of the strip. 
It turns out that the linear dilaton has, once again, the status of a critical case between two qualitatively different regimes.~\footnote{
We  notice that the large $d$ limit of \eqref{eq:nuprime} implies $\bar\nu\to 1$. In this limit the minimal surfaces tend to behave effectively as if they were in the linear dilaton background. 
A related phenomenon  pointed out in \cite{Jarvinen:2024vdi} is that 
 the linear dilaton background appears  from general relativity  when taking the limit of large number of compactified dimensions.
 }
The numerical results for the HEE in $\MnuM$ assuming $d=2$ are presented in Fig.\,\ref{fig:HEE_MnuM}.  A similar qualitative behavior happens for other values of $d$.

For $\nu < 1$ (and $\nu=0 $ if $d>2$), we obtain that $S_A$ reaches a plateau at large $\L_A$, with value
\be S^{\rm max}_A = \frac{A_{d-2}}{2 G_N \eta (\nu^2 + d -2)}\left(\frac{r_b}{L} \right)^{d-2}\,.
\label{eq:Smax}
\ee 
The existence of HEE plateaux has been noted in the literature, typically at finite temperature, see \cite{Hubeny:2013gta}.  Here we observe this saturation phenomenon at \textit{zero temperature}. 
For $d=2$ the plateau occurs  for $\nu>0$, while for  $d>2$  the plateau occurs  for any $\nu\geq 0 $, hence even for AdS. 
From the viewpoint of the general metric \eqref{eq:ds2_gen}, a HEE plateau occurs when the integral $\int_{0}^{r_b} dr a(r)^{\frac{d-1}{2}} b(r)$ is finite.

For $\nu=1$, we know that the base of the smooth surface is bounded from above. This follows from the direct calculations of the area  in section~\ref{se:surfaces_Mnu}, and from the application of condition~\eqref{eq:HEEcond2} from the viewpoint of the general warped metric: 
the upper bound is {$\frac{\pi L}{\eta r_b (d-1)}$}. There is thus necessarily a phase transition, which happens to occur precisely at this value. Using the explicit formula for the linear dilaton geodesic length \eqref{eq:Area_LD}, we find that $S^{\nu=1}_A$ experiences a discontinuity in its second derivative. Therefore there is a \textit{second order} phase transition in the HEE.

For $\nu>1$  we observe  that the base of the geodesics is bounded and that a doubling of the geodesics occurs. From the viewpoint of the general metric, we understand that this phenomenon is a consequence 
of the curvature singularity that truncates space, see section \ref{se:PT_boundary}. 
The phase transition occurs when the area of the smaller smooth surface crosses the area of the square surface. We find that this occurs approximately at {$L_A \simeq \frac{2 L}{\eta(\nu^2 + d-2)r_b}$}.\,\footnote{
Notice that this condition depends on the $\frac{\eta r_b L_A}{L}$ combination. This implies that an alternative way to reach the saturation behavior described in \eqref{eq:Smax} is to increase $r_b$ at fixed $L_A$. Namely, for $\nu<1$, taking large $r_b$ implies $r_b\gg r_0$, which is the  limit computed  in \eqref{eq:Deltas_r0small}.  For $\nu>1$, taking large $r_b$ implies  that the square surface necessarily dominates since $L_A \gg \frac{2L}{\eta(\nu^2 + d-2) r_b}$ in that limit. Therefore in both cases we have  $S=S_A^{\rm max}$.    
} 
The first derivative is discontinuous, hence the HEE experiences a \textit{first order} phase transition.  
In short: 
\begin{mdframed}[style=EqFrame]
\begin{equation}
\shortstack[l]{
\textrm{A curvature singularity at finite conformal distance implies} \\ \textrm{a HEE first order phase transition.}}
\end{equation}
\end{mdframed}
{The HEE behavior,  computed here  from the bottom up,  closely resembles the one obtained from stringy setups in \cite{Klebanov:2007ws,Jokela:2020wgs,Jokela:2023lvr,Fatemiabhari:2024aua}. }

\section{Stability and the Radion Effective Action}
\label{sec:radion}

In this section and the next, we study the brane-dilaton system globally. To this end, we put the classical bulk solutions (see~\cite{Fichet:2023dju}) into the fundamental action $\cal S$, defined in \eqref{eq:action}.   
This defines a ``holographic'' on-shell action that depends only  on the brane location $r_b$ and the value of the dilaton on the brane, $\phi_b$, 
${\cal S}_{\rm on-shell}[r_b,\phi_b]$.  We restrict the spacetime dimension to $D=5$ for simplicity. We remind that we restrict to static configurations, with   $r_b$  independent on the spacetime coordinates $x^\mu$, see section~\ref{subsec:Mnu_spacetime}.

\subsection{The  Effective Potential}

\label{se:Effective_Potential}

At zero-th order in the fluctuations, the on-shell action  corresponds to an effective potential  ${\cal S}_{\rm on-shell}\equiv -\int d^4x \, V_{\rm eff}(r_b,\phi_b) $. 
In~\cite{Fichet:2023dju} we found 
\be
V_{\rm eff} = U_b(\phi_b)\left(\frac{r_b}{L}\right)^4\,,\quad \quad \quad 
U_b(\phi_b) = V_b(\phi_b) +\Lambda_b \mp 6M_5^3\eta(\phi_b)   \,,
\label{eq:Veff}
\ee
for ${\cal M}^\mp_\nu$. We introduced the dilaton-dependent scale  $\eta(\phi_b)= ke^{\nu\bar\phi_b}$, with $\eta(v_b)\equiv  \eta$ at the minimum $\phi_b\equiv v_b$.

Stability along the $r_b$ direction requires to tune the 4D cosmological constant to {$\Lambda_b=\pm 6M_5^3\eta$} for ${\cal M}^\mp_\nu$. 
Stability along the $\phi_b$ direction requires $U_b''>0$, which translates as a condition on $V_b$ \cite{Fichet:2023dju}. In the $\MnuM$ spacetime such  a vacuum is only metastable since $V_{\rm eff}$ is unbounded from below at large $\phi_b$. We will see in section~\ref{se:PT} that this feature is exacerbated in the presence of a very large black hole.

\subsection{The Radion}

Both the dilaton and spacetime can fluctuate around the configuration determined by $(r_b,v_b)$. 
Upon gauge fixing, these  scalar fluctuations reduce to a single degree of freedom, usually called the \textit{radion} field (see~\cite{Csaki:2000zn,Fichet:2023dju}).  It was shown in~\cite{Fichet:2023dju} that the radion spectrum in $\Mnu^\pm$ always contains an isolated mode $R_0(x)$, that we  refer to as \textit{the} radion mode. The rest of the radion spectrum may be either discrete or continuous depending on $\nu$, see summary table~\ref{tab:summary}. Notice that this is tied to the singularity being either at finite or infinite conformal distance. 

\subsubsection{Definition}

We consider the fluctuation of the dilaton and of the scalar sector of the fluctuations of the metric. In conformal coordinates, upon gauge-fixing (see \cite{Csaki:2000zn,Fichet:2023dju}) the following parametrization can be used: 
\begin{align}
\nn ds^2&=  e^{-2 A(z)} \left[ e^{-2F(x,z)} \eta_{\mu\nu} dx^\mu dx^\nu+(1 + 2 F(x,z))^2 dz^2\right]  \,, \label{eq:ds2fluc}
\\
\bar \Phi(x,z) &= \bar \phi(z) +\bar\varphi(x,z)  \,.
\end{align}
It was shown in \cite{Fichet:2023dju} that the equations of motion relate  the dilaton fluctuation $\varphi(x,z) $ to the  $F(x,z) $ fluctuation by 
\begin{equation}
\bar\phi^\prime(z) \bar\varphi(x,z) = \left( \partial_z  - 2 A^\prime(z) \right) F(x,z) \,. \label{eq:EoMvarphi}
\end{equation}

We introduce a spectral decomposition over an orthogonal basis of modes
\begin{equation}
  F(x,z)= \sum_{\lambda} F_\lambda(z)R_\lambda(x)  \,, \qquad  \textrm{and} \qquad \varphi(x,z)= \sum_{\lambda} \varphi_\lambda(z)R_\lambda(x)   \,,   \label{eq:FR}
\end{equation}
where each profile $F_\lambda(z)$ satisfies the equation of motion with mass $p^2=-m_\lambda^2$. We use  $\lambda=0$ to label the isolated radion mode of our interest, with profile   
 $F_0(z)$ and mass $m_0$. The summation over modes in Eq.~(\ref{eq:FR})   may be either discrete or continuous~\cite{Fichet:2023dju}.

\subsubsection{Radion profile}

{The fluctuation in mixed position-momentum space is defined by the 4D Fourier transform $F(x^\mu,z)= \int \frac{d^4 p}{(2\pi)^4} e^{i p_\mu x^\mu} F(p,z)$}.   The equation of motion for the profile $F_\lambda(z)$ is \cite{Csaki:2000zn,Megias:2021mgj, Fichet:2023dju}  
\begin{equation}
\partial_z^2 F_\lambda(z) + \frac{1 + 2 \nu^2}{\nu^2 - 1} \frac{1}{z} \partial_z F_\lambda(z) + m_\lambda^2 F_\lambda(z) = 0 \,.  \label{eq:EoMF}
\end{equation}
 $F_\lambda(z)$ satisfies the following boundary conditions:
\begin{itemize}
\item[{\it i)}] Regularity at $z=0$ if $z \in (0,z_b]$, or at $z \to \infty$ if $z \in [z_b,\infty)$.
\item[{\it ii)}] Boundary condition at the brane~\cite{Fichet:2023dju}:
\be
\left( \partial_z - 2 A^\prime(z) \mp \frac{2 m_\lambda^2 e^A}{U_b^{\prime\prime}}  \right) F_\lambda(z) \bigg|_{z = z_b}  = 0 \,,  \label{eq:BC2}
\ee
where the $-$ sign  stands for $z \in (0,z_b]$, i.e.~for $\mathcal M^+_{\nu<1}$ and $\mathcal M^-_{\nu > 1}$, and the $+$ sign stands for $[z_b,\infty)$, i.e.~for $\mathcal M^-_{\nu<1}$ and $\mathcal M^+_{\nu > 1}$.  This boundary condition is generated by the brane terms in the action~(\ref{eq:action}), as discussed in e.g.~Ref.~\cite{Csaki:2000zn}.
\end{itemize}

Eq.~(\ref{eq:BC2}) constitutes the eigenvalue equation whose solution
for the lightest mode $m_0$ corresponds to the radion mass. 
Solving for a small mass $m_0\ll \eta$ we find the profiles
corresponding to the radion mode $R_0$,
\begin{equation}
F_0(z) \simeq \left\{ 
\begin{array}{cc}
C_F \left( \frac{z}{z_b} \right)^{2\gamma}  & \qquad {\rm in}\quad\mathcal M_\nu^+  \\
C_F \left( 1 + \frac{m_0^2}{4(\gamma - 1)}  z^2\right) & \qquad  {\rm in}\quad \mathcal M_\nu^- 
\end{array}   \right. 
 \,,     \label{eq:F0}
\end{equation}
and
\begin{equation}
\varphi_0(z) \simeq \left\{ 
\begin{array}{cc}
\sqrt{3 M_5^3} C_F \nu \left( \frac{z}{z_b}\right)^{2\gamma}  & \qquad {\rm in}\quad \mathcal M_\nu^+  \\
-2 \sqrt{ 3 M_5^3} \frac{C_F}{\nu} \left(  1 - \frac{\nu^2-1}{6} m_0^2 z^2 \right)  & \qquad {\rm in}\quad \mathcal M_\nu^- 
\end{array}   \right.   \,.
\label{eq:varphi}
\end{equation}

The normalization constant is determined by 
substituting the solutions in the on-shell action and 
requiring canonical normalization,
\begin{equation}
S_{R_0,{\rm kin}} = -\int d^4x dz \, e^{-3A(z)}\left( 6 M_5^3 |F_0(z)|^2 +  |\varphi_0(z)|^2   \right) (\partial_\mu R_0)^2  \equiv  - \frac{1}{2} \int d^4x  (\partial_\mu R_0)^2 \,,
\end{equation}
which sets
\begin{equation}
C_F = \left\{ 
\begin{array}{cc}
\frac{|1 - \nu^2|}{\sqrt{2}} \frac{ \eta^{3/2} z_b }{\sqrt{3 M_5^3}} & \qquad {\rm in}\quad    \mathcal M_\nu^+
 \\
\frac{\nu |1 - \nu^2|}{2} \frac{\eta^{3/2} z_b}{\sqrt{3 M_5^3}}   & \qquad {\rm in}\quad  \mathcal M_\nu^- 
  \end{array} \right. \,.
  \label{eq:CF}
\end{equation}
{While the determination of $C_F$ is relevant for  canonical normalization,  it is of course not necessary  for calculating the radion mass.}

\subsection{Radion Mass from the On-Shell Action}

 We show how to compute the radion mass directly from the on-shell action.
This provides an important consistency check of the equation of motion and boundary condition
\eqref{eq:EoMF},\,\eqref{eq:BC2}, from which the mass can be independently derived. This is also a necessary step to further derive the radion effective action, including its interactions with other quantum fields.  {The full radion quadratic action is}
\begin{equation}
\mathcal S_{R_0} =    - \int d^4x  \left[ \frac{1}{2} (\partial_\mu R_0(x))^2 + \frac{1}{2} m_0^2 R^2_0(x)  \right] \,.  \label{eq:S_R0}
\end{equation}
{To obtain the radion mass term  we first rewrite the full renormalized on-shell action as} 
\begin{equation}
\S_{\rm on-shell} = \S_{\rm bulk}^{\rm ren} + \S_{\rm brane} + \S_{\rm GHY}  \,,  \label{eq:S_onshell}
\end{equation}
with
\begin{eqnarray}
\S_{\rm bulk}^{\rm ren} &=& \int d^5x  \sqrt{g}  \bigg(  \frac{M_5^3}{2} {}^{(5)}R  - \frac{1}{2} (\partial_M \phi)^2 - V(\phi)  \bigg) + \S_{\rm ct} \,,\\ 
\S_{\rm brane} &=& -\int_{\textrm{brane}} d^4x \sqrt{\bar g} \left(V_b(\phi_b)+\Lambda_b\right)   \,, \\
\S_{\rm GHY} &=& M_5^3 \int_{\textrm{brane}} d^4x \sqrt{\bar g} \, K \,.
\end{eqnarray}
{The integral in $\mathcal M^+_\nu$ is infinite. We regularize it with a}  counterterm given by~\cite{Mann:2009id,Fichet:2023dju}
\begin{equation}
\S_{\rm ct} =  2 M_5^3 \int_\Sigma d^4x \sqrt{h} \, \eta(\phi_\Sigma) \,, 
\end{equation}
with $\Sigma$ a 4D cutoff surface located at $r_\Sigma \gg r_b$, and $h_{\mu\nu}$ being the induced metric on $\Sigma$. {The counterterm receives contributions from the bulk, and also from the boundary due to the GHY term at $r_\Sigma$.}

We expand the action at quadratic order 
in the radion field $R(x)$. Notably, the expansion of the metric determinant is 
\begin{eqnarray}
g &\equiv& |\det g_{MN}| = e^{-5 A(z)} \left( 1 - 2 F(x,z) - \frac{5}{2} F^2(x,z) + \cdots \right)  \,.
\end{eqnarray}
Focussing on the radion mode $R_0$, 
we substitute the expressions of the $F_0$, $\varphi_0$ profiles \eqref{eq:F0} and \eqref{eq:varphi} in the renormalized on-shell action.   We find the following contributions  to the $R_0^2$ term:
\begin{eqnarray}
\S_{\rm bulk, mass}^{\rm ren} &=& \left\{
\begin{array}{cc}
  \frac{16 - 24 \nu^2 - 3 \nu^4}{6 (1-\nu^2)^2} \frac{1}{z_b^2} \int d^4x R^2_0(x)  &  \qquad  (\mathcal M_\nu^+)    \\ 
 - \frac{13}{3} \frac{\nu^2}{(1-\nu^2)^2} \frac{1}{z_b^2} \int d^4x R^2_0(x)  &  \qquad  (\mathcal M_\nu^-)    
   \end{array} \right.  \,,  \\
\S_{\rm brane, mass} &=& \left\{
\begin{array}{cc}
  - \frac{U_b^{\prime\prime} \nu^2 - 2 \eta (4 - \nu^2)^2}{4 (1-\nu^2)^2} \frac{1}{\eta z_b^2} \int d^4x R^2_0(x)  &  \qquad  (\mathcal M_\nu^+)    \\ 
- \frac{U_b^{\prime\prime} + 18 \nu^2 \eta }{ 2 (1- \nu^2)^2} \frac{1}{\eta z_b^2} \int d^4x R^2_0(x)  &  \qquad  (\mathcal M_\nu^-)
   \end{array} \right.  \,, \\
\S_{\rm GHY, mass} &=& \left\{
\begin{array}{cc}
  - \frac{8}{3}\frac{(4 - 3 \nu^2)}{(1-\nu^2)^2} \frac{1}{z_b^2} \int d^4x R^2_0(x) &  \qquad  (\mathcal M_\nu^+)   \\ 
 \frac{40}{3} \frac{\nu^2}{(1-\nu^2)^2} \frac{1}{z_b^2} \int d^4x R^2_0(x)  &  \qquad  (\mathcal M_\nu^-)
   \end{array} \right.   \,.
\end{eqnarray}
We have made use of the following identities
\begin{equation}
V_b(v_b) = 0 \,, \quad V_b^\prime(v_b) = \pm 2 \sqrt{3 M_5^3} \nu \eta \,, \quad V_b^{\prime\prime}(v_b) = U_b^{\prime\prime}(v_b) \pm 2 \nu^2 \eta \,, \quad \Lambda_b = \pm 6 M_5^3 \eta \,,
\end{equation}
for $\mathcal M_\nu^{\mp}$. The first condition is a choice with no loss of generality. The second and third conditions follow from the definition of the brane-localized effective potential $U_b(\phi_b)$ {given} in \eqref{eq:Veff}, using that $U_b$ has a minimum at $\phi_b = v_b$. The fourth condition corresponds to the tuning that ensures  $U_b(v_b) = 0$, so that the effective radion potential $V_{\rm eff}(r_b,v_b) = U_b(v_b) \left( \frac{r_b}{L} \right)^4$ has a flat direction in $r_b$.

The full mass term is then 
\begin{equation}
\S_{R_0,{\rm mass}} = \left\{
\begin{array}{cc}
  - \frac{\nu^2}{4(1-\nu^2)^2} U_b^{\prime\prime}  \frac{1}{\eta z_b^2} \int d^4x R^2_0(x)  &  \qquad  (\mathcal M_\nu^+)   \\ 
  - \frac{1}{2(1-\nu^2)^2} U_b^{\prime\prime}  \frac{1}{\eta z_b^2} \int d^4x R^2_0(x)  &  \qquad  (\mathcal M_\nu^-) 
   \end{array} \right.  \,.
\end{equation}
From {comparison with Eq.~(\ref{eq:S_R0})}, we identify the radion mass
\begin{equation}
m_0^2 = \left\{
\begin{array}{cc}
\frac{\nu^2}{2(1-\nu^2)^2} U_b^{\prime\prime}  \frac{1}{\eta z_b^2}  &  \qquad  (\mathcal M_\nu^+)    \\ 
\frac{1}{(1-\nu^2)^2} U_b^{\prime\prime}  \frac{1}{\eta z_b^2}  &  \qquad  (\mathcal M_\nu^-)   
\end{array}  \right.  \,.
\end{equation}
This result agrees with the pole mass obtained from the radion propagator~\cite{Megias:2021arn,Megias:2023kpk,Fichet:2023dju} 
\begin{equation}
i G_F(z,z^\prime; p^2) = \sum_{\lambda} \frac{F_\lambda(z) F_\lambda^\ast(z^\prime)}{p^2 + m_\lambda^2 - i \varepsilon} \,
\end{equation}
in the limit of small $U_b^{\prime\prime}$. We have similarly checked the mass term at next-to-leading order in the $U_b^{\prime\prime}$ expansion, finding again consistency with the result from \cite{Fichet:2023dju}.

\section{Holographic    Thermal Phase Transitions }
\label{sec:PT}

We turn to the on-shell action in the presence of the bulk black hole. 
As  discussed in Ref.~\cite{Fichet:2023dju}, the black hole is always in the $\Mnu^-$ region. 
We use the $(x^\mu,r)$ coordinates with $r\in (0,r_b]$. Throughout this section we use a $ Z_2$ orbifold convention as in \cite{Shiromizu:1999wj}
which implies that the spacetime is mirrored on the other side of the brane. The domain of integration of $r$ is thus doubled in the action. A notable implication is that the entropy of the black hole horizon is doubled when using this convention, see {e.g.}~Ref.~\cite{Megias:2018sxv}.

The black hole metric is given in \eqref{eq:ds2_Mnu_BH}.
We assume that the brane is outside the black hole,  $r_h < r_b$.
We define $f(r_b)\equiv f_b$, with $f(r)=1-\left(\frac{r_h}{r}\right)^{4-\nu^2}$. 
We introduce the brane proper time $dt = \frac{r_b}{L} \sqrt{f(r_b)} d\tau$, with $\tau$ the comoving time from the comoving volume element $d^4x = d\tau d^3x$.

We find the on-shell action   
\be
{\cal S}_{\rm on-shell}=  -\int dtd^3x \left( \frac{r_b}{L} \right)^3\, U_b(\phi_b)  \,,\quad \quad 
\label{eq:Veff_rh} U_b(\phi_b)=U^0_b(\phi_b) +6M_5^3 \eta(\phi_b) \u(r_b,r_h) \,,
\ee 
with 
\be
\u(r_b,r_h) = \frac{1}{\sqrt{f_b}} \left( \sqrt{f_b} - 1 + \frac{(2+ \nu^2)}{6} (1 - f_b) \right) \label{eq:Xdef} \,.
\ee
The $U^0_b(\phi_b)$ term in \eqref{eq:Veff_rh} is $U^0_b(\phi_b)=U_b(\phi_b)|_{r_h\to0}$, i.e.~it is the brane-localized effective potential with no black hole given in \eqref{eq:Veff}. The effect of the black hole is fully encoded into the second term. 

\subsection{Thermodynamics on the Brane}

From the viewpoint of the brane, the black hole contributes as a \textit{perfect fluid}, see \cite{Fichet:2022ixi, Fichet:2023dju, Fichet:2022xol, Fichet:2023xbu}. It was shown in \cite{Fichet:2023dju}
 that the thermodynamics of this holographic fluid essentially mirrors the bulk black hole thermodynamics. 

\subsubsection*{Volume}

A first thermodynamic  variable defined on the brane
is the volume $(V_b)$, which is related to the comoving volume $V_3 = \int d^3x$ by
\be 
V_b \equiv V_3 \, a^3(r_b) \,,
\ee 
with  the scale factor
\begin{equation}
a(r_b) \equiv e^{-A(r_b)} = \frac{r_b}{L} \,. 
\end{equation}

\subsubsection*{Temperature}

 The fluid temperature $(T_b)$ is related to the Hawking temperature $T_h$ as~\cite{Fichet:2023dju}
\begin{equation}
T_b = \frac{T_h}{a(r_b)} = \frac{4 - \nu^2}{4} \frac{\eta}{\pi} \left(  \frac{r_b}{r_h} \right)^{\nu^2- 1} \,.
\end{equation}
 The Hawking temperature $T_h$ is obtained by requiring the absence of conical singularity near the horizon (see e.g. \cite{Fichet:2022xol} for the general planar black hole case). 

We can notice that $T_b$ {decreases (increases) with $r_b/r_h$ for $\nu<1$ ($\nu>1$). For $\nu=1$, the fluid has a constant temperature {$T_b|_{\nu = 1} =\frac{3\eta}{4\pi}$} whenever it exists.

\subsubsection*{Free energy}

The on-shell action can be expressed in Euclidean time $(t = -i t_E)$ as $i {\cal S}_{\textrm{on-shell}} =  - {\cal S}^E_{\textrm{on-shell}}$. 
When $\phi_b$ is at its vev {$v_b^T$},  we identify the free energy of the system:~\footnote{{Notice that in the presence of a black hole, the vev $(v_b^T)$ is different from the vev with no  black hole $(v_b)$, and thus $U_b^0 \equiv U_b^0(v_b^T)$ is generically non-vanishing in the former case. Nevertheless, this effect is small for $r_b$ large enough compared to $r_h$, as  can be seen in Fig.~\ref{fig:Ub}.}} 
\begin{equation}
{\cal S}^E_{\textrm{on-shell}}\bigg|_{\phi_b=v_b}    \equiv \beta V_b { F} \,.
\label{eq:free_energy_def0}
\end{equation}
Here ${ F}$ is the free energy density and $\beta=\frac{1}{T}$ the inverse temperature of the system. 

At equilibrium, we have $T=T_b$ as imposed by the  absence of conical singularity upon Euclidean time compactification. 
In this section we will briefly discuss the out-of-equilibrium case of $T\neq T_b$, in which  case the free energy has an extra contribution $F_{\rm cone}\propto (T-T_b)$ \cite{Creminelli:2001th,Megias:2020vek}. The $T_b$ temperature can be identified even in case of non-vanishing conical singularity, see \cite{Creminelli:2001th}. 

The free energy of the system is given by
\begin{equation}
 { F} = U_b^0 + {F}_{\rm fluid}+F_{\rm cone}\,,\quad\quad 
{ F}_{\rm fluid}= 6M_5^3 \eta \Delta(T_b) \,,
\label{eq:free_energy_def}
\end{equation}
 with $\Delta(T_b)\equiv \Delta(r_b,r_h)$. 
In the small black hole limit $r_h\ll r_b$, the fluid free energy {density} simplifies to 
\be
F_{\rm fluid}\bigg|_{r_h\ll r_b}\approx -(1-\nu^2 ) \eta M_5^3 \left( \frac{r_h}{r_b}\right)^{4 - \nu^2} \,.\label{eq:F_small_rh}
\ee
This reproduces the result presented in \cite{Fichet:2023dju}. In this limit the free energy is negative for $\nu<1$ and positive for $\nu>1$. We will see in section \ref{se:PT} that this is not true away from the small $r_h$ limit, which will be the cause of a phase transition. In contrast,  \eqref{eq:Xdef} implies that the vanishing of $F_{\rm fluid}$ for $\nu=1$ remains exact for any $r_h$.

A typical out-of-equilibrium situation is when a thermal bath with temperature $T$ is localized on the brane, while the black hole/holographic fluid has $T_b\ll T$. 
The thermal bath radiates into the bulk, feeding the out-of-equilibrium black hole. In such a setup the bulk is non-empty, and a Vaidya-type metric has to be used \cite{Langlois:2002ke,Langlois:2003zb}.  Here we will discuss this case only qualitatively.

\subsection{Phase Transitions}

\label{se:PT}

The information about phase transitions is encoded into the $\Delta(T_b)$ function. A phase with no fluid, i.e.~with no black hole ($r_h\to 0$), has $f_b|_{r_h\to 0} = 1$ and thus {$\Delta(T_b)|_{r_h\to 0} = 0$}. 
The existence of phase transitions is thus controlled by the sign of $\Delta(T_b)$. 

\paragraph{Critical temperature.} The critical value of $T_b$ at which a phase transition occurs is denoted {by} $T_c$ and is determined by $\Delta(T_c)=0$. Written in terms of $f_b$, the solutions
are given by \be \sqrt{f_b}_\pm = \frac{3\pm|\nu^2-1|}{2+\nu^2}\,. \ee
We also know that physical solutions must satisfy $f_b\leq 1$. 

\paragraph{Latent heat.}
The order of the phase transition is controlled by the associated latent heat per unit volume, {$\ell_b$}. The transition is first order if {$\ell_b \neq 0$}   and second order if {$\ell_b = 0$}. The latent heat for a phase transition at temperature $T_c$ is defined as {$L_b = T_c\Delta S$}. The entropy in each phase satisfies {$S = - \partial (V_b F) /\partial T_b|$}. {If $r_b$ is large enough compared to $r_h$ (typically $r_b \gtrsim 1.1 r_h$) then $v_b^T \simeq v_b$, and} the difference between the two phases is only due to the fluid free energy, which furthermore vanishes identically in the phase with no black hole. 
Putting the pieces together and using the chain rule, we derive the latent heat for the transition from the no black hole phase to the  black hole phase  at a given critical temperature $T_c$ via
\be
\ell_b = - T_c  \frac{\partial F_{\rm fluid}}{\partial r_b}\frac{\partial r_b}{\partial T_b}\bigg|_{r_{b,c}}\,.
\ee

\subsubsection*{Case $\nu<1$}

For $\nu\in[0,1)$, we have $f_-=1$ while the other solution gives $f_+>1$ for any value of $\nu$, which is not physical. 
Therefore $\Delta(T_b)$ cannot change sign as a function of $T_b$. It satisfies {$\Delta(T_b) \leq 0$} for all values of $T_b$, with $\Delta(T_b) \to 0$ in the limit of no black hole $r_h \to0$, for which $T_b\to 0$. 

We conclude that the black hole phase is energetically favored whenever it exists. The critical temperature is thus $T_c=0^+$. The latent heat vanishes, hence it is a \textit{second order} phase transition. 

The $F(T_b)$ curve is pictured in Fig.\,\ref{fig:F_curves}, left panel.

\subsubsection*{Case $\nu>1$}

For $\nu\in(1,2)$, we have $f_+=1$ and 
\be
\sqrt{f}_- = \frac{4-\nu^2}{\nu^2+2}\,,
\ee
which  is  inside the $[0,1]$ interval. 
There is thus a phase transition at a nonzero value of $r_h/r_b$, 
 \be
\frac{r_h}{r_{b}}\bigg|_c= \left(\frac{12(\nu^2-1)}{(\nu^2+2)^2}\right)^{\frac{1}{4-\nu^2}}\,.
\label{eq:r_critical}
\ee
The corresponding critical  temperature is 
\be
T_b = \frac{4-\nu^2}{4}\frac{\eta}{\pi}\left(\frac{(\nu^2+2)^2}{12(\nu^2-1)}\right)^{\frac{\nu^2-1}{4-\nu^2}}\equiv T_c\,.
\ee
In terms of the horizon position, it turns out that {$\Delta(T_b) > 0$} for $r_h<r_{h,c}$ and {$\Delta(T_b) < 0$} for $r_h>r_{h,c}$.

We find a latent heat per unit volume that is finite  for any $1<\nu<2$,
\be
 {\ell_b} = -\frac{12(\nu^2-1)}{4-\nu^2}M_5^3\eta \,,
\ee
therefore the phase transition is \textit{first order}.  

Furthermore, the latent heat is \textit{negative}. This is tied to the fact that for $\nu>1$, $T_b$ decreases when $r_b$ approaches  $r_h$.
This implies that there is a cusp in the free energy curve. This is shown in Fig. \ref{fig:F_curves}, right panel.

\begin{figure}[t]
\centering
\includegraphics[trim={5cm 5cm 9.5cm 7cm},clip,width=0.35\textwidth]{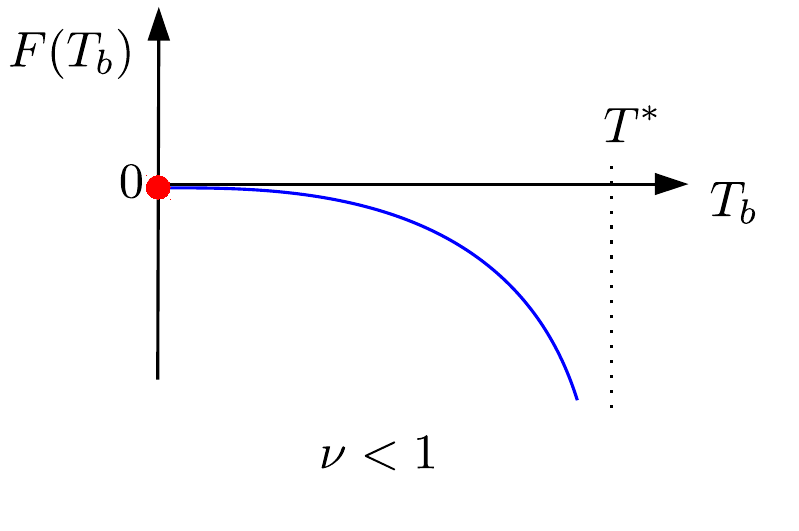}
\includegraphics[trim={7.4cm 5cm 9.5cm 7cm},clip,width=0.29\textwidth]{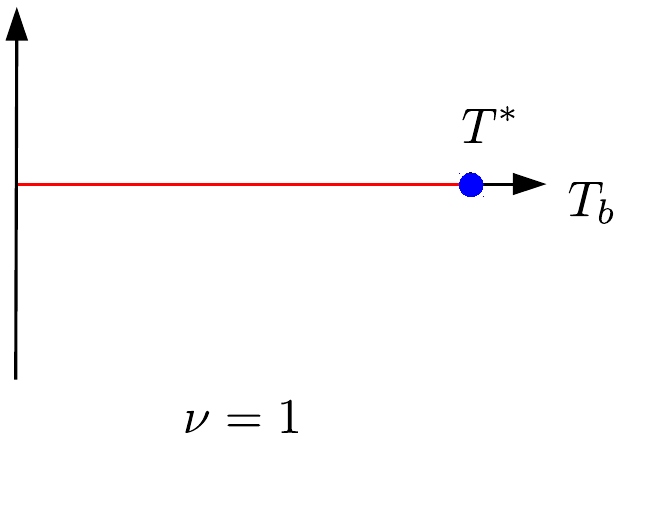}
\includegraphics[trim={7.4cm 5cm 7.7cm 7cm},clip,width=0.34\textwidth]{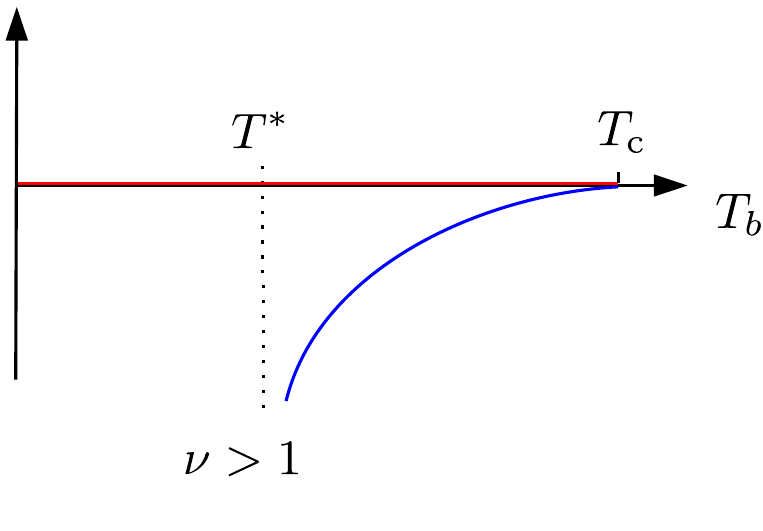}
\caption{\it    Free energy curves and thermal phases as a function of the temperature of the holographic fluid.  Red  corresponds to the phase with no black hole, blue  corresponds to the  black hole phase.  The black hole  phase is stable at any value of $T_b$ shown on the graph.
}
\label{fig:F_curves}
\end{figure}

\subsubsection*{Case $\nu=1$}

For $\nu=1$ i.e.~the linear dilaton background,  the fluid has constant temperature  $T_b=\frac{3\eta}{4\pi}$ and vanishing free energy,  $F_{\rm fluid}=0$, as shown in \cite{Fichet:2023xbu,Fichet:2023dju}. That is, the fluid has Hagedorn behavior. 
We conclude that the phase transition occurs at $T_c=\frac{3\eta}{4\pi}$ and is of second order. This is shown in Fig. \ref{fig:F_curves}, middle panel. 

\subsection{Discussion}

\subsubsection*{Heat Capacity} We computed the heat capacity at constant volume of the black hole, which  is given by
{$C_V = \left( \partial E / \partial T_b \right)_{V_b}$}, where $E$ is
the energy of the system. We find that the heat capacity is positive
at all values for which the black hole phase is energetically
favored. Hence the black hole is thermodynamically stable for all
values of $\nu$.

\subsubsection*{Temperature bound}
For $\nu<1$ and $\nu>1$  the fluid free energy tends to negative infinity when the black hole approaches the brane. In this limit
the temperature approaches the finite value 
\be
T^*= \frac{4-\nu^2}{4\pi}\eta\,.
\ee
For $\nu<1$, $T^*$ is approached from below, i.e.~the fluid temperature is constrained to the interval  $T_b\in[0 , T^*  )$.
For $\nu>1$, $T^*$ is instead approached from above, i.e.~the fluid temperature is bounded from below, $T_b\in ( T^* , \infty )$. 
Notice that in the $\nu=1$ case, the temperature is restricted to the single value $T^*$ {(see also \cite{Gursoy:2010jh,
Elander:2020rgv})}.

\subsubsection*{Cosmological scenario}

The thermodynamic behavior of the black hole is perhaps more intuitively understood when considering a brane cosmology--type scenario where a thermal bath of temperature $T\gg T_b$ is present on the brane. The thermal bath leaks energy into the bulk, feeding the bulk black hole, which is assumed to not exist initially (see~\cite{Gubser:1998bc,Langlois:2002ke,Langlois:2003zb} for  analyses of such scenarios in AdS). The rigorous evolution of such cosmological scenarios is done using a Vaidya-type metric. Here we discuss them only qualitatively to get a sense of the black hole behavior.  

For $\nu<1$ and $\nu=1$, the black hole is created via a second order {phase} transition whenever some energy leaks into the bulk. For $\nu<1$ the temperature grows while the black hole size increases. For $\nu=1$ the temperature is instead constant and non zero for any black hole size, since it is independent of $r_h$. 

For $\nu>1$, the thermal bath should heat up the bulk sufficiently such that  a first order {phase} transition occurs and the black hole directly forms with finite size given by \eqref{eq:r_critical}.\,\footnote{The first order phase transition may proceed via bubble nucleation and thus produces gravitational waves, see e.g.~\cite{Creminelli:2001th,Megias:2018sxv,Baratella:2018pxi,Lee:2021wau,Megias:2023kiy,Mishra:2024ehr}.}
The black hole temperature  decreases is its size increases further.

\subsection{Instability  from Big Black Holes }

\begin{figure*}[t]
\centering\includegraphics[width=0.43\textwidth]{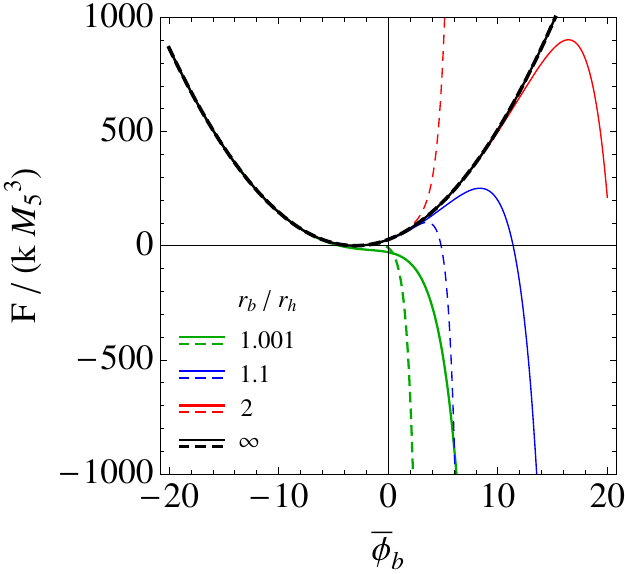}  
 \caption{\it The total free energy density  as a function of $\bar\phi_b$, assuming the brane-localized effective potential $V_b(\phi_b)=\frac{\gamma}{2}(\phi_b-v_b)^2$. Solid (dashed) lines are for $\nu = 0.5\,(1.5)$.  We have considered $\bar v_b = -3$ and $\gamma = 2k$. Unstability of the system occurs when the horizon comes close to the brane.}
\label{fig:Ub}
\end{figure*}

We study the stability of the brane-dilaton system in the presence of the black hole. This is described by the brane-localized effective potential $U_b(\phi_b)$ given in \eqref{eq:Veff_rh}. Notice {that} this is equivalent   to study the free energy at $\phi_b\neq v_b$. 

The stabilization of $\phi_b$ is driven by the brane potential $V_b(\phi_b)$, see \eqref{eq:Veff}. 
At zero temperature, due to the $\phi_b$ dependence of $\eta$, 
requiring $U''_b(v_b)>0$ implies the stability condition $V_b''(v_b)>2\nu^2\eta$ \cite{Fichet:2023dju}. 

At finite temperature, the extra contribution from the black hole  in  \eqref{eq:free_energy_def} is also proportional to $\eta(\phi_b)$. In the black hole phase this term is  negative, since by definition we have $\Delta(r_b,r_h)<0$ for the black hole phase to be energetically favored. 

When the black hole becomes so big that its horizon approaches the brane, $r_h\sim r_b$, the effect of the black hole dominates and we have $\Delta(r_b,r_h)\sim -\frac{4-\nu^2}{6\sqrt{f_b}} $ with $f_b$ approaching zero. 
Therefore this black hole term tends to \textit{destabilize} the effective potential. 
The critical value for $r_h/r_b$ depends on the steepness of the $U_b$ potential, {and}  is approximately given by
\be
\frac{r_h}{r_b} \approx  1- \frac{\nu^4}{9}  \frac{4-\nu^2}{\left(U_b^{0\,\prime\prime}(v_b^T)\right)^2} \eta^2(v_b^T) \,.
\ee
Notice that in the stiff potential limit {$U_b^{0\,\prime\prime} \to \infty$}, the horizon can reach the brane since the potential cannot be destabilized.  

We conclude that, for finite $U_b^{0\,\prime\prime}$, the $T^*$ temperature cannot be actually reached. A radical change in the system occurs for a value of $r_h$ slightly below $r_b$: the classical effective potential for $\phi_b$ becomes unbounded from below --- and can only get stabilized at loop level.
Numerical examples of this effect are presented in Fig.~\ref{fig:Ub}, where we used $V_b(\phi_b)=\frac{\gamma}{2}(\phi_b-v_b)^2$.

\section{Summary}
\label{se:conclusion}

In this paper we studied  holographic entanglement entropy  and revisited thermal {phase transitions} and confinement in dilaton gravity, with special emphasis on their interplay with bulk curvature singularities.

We  mostly focus on  a very simple class of backgrounds {$\mathcal M_\nu$,  parameterized by a real parameter $\nu$ interpolating  AdS ($\nu=0$) and  linear dilaton ($\nu=1$)}, that retains the essence of the phenomena we want to highlight. We also prove some properties in general warped metrics.
The considered backgrounds are non-asymptotically AdS and feature a flat brane on which the holographic theory is defined. 
Curvature singularities can be either at infinite or finite conformal distance from the brane. 
We show that the latter case is automatically implied by a metric zero at finite conformal distance, in which case the singularity acts as a regular boundary that truncates spacetime. 

We show that the behaviors of  the entanglement, thermodynamics, and confinement {properties} are all tied to the nature of the bulk singularity.  Some important results are summarized in Tab.\,\ref{tab:summary}.

 \begin{table}[t]
    \centering
    \begin{tabular}{|c|c|c|c|c|c|c|c|}
    \hline
       Value of $\nu$  & $0$ [AdS$$] & $(0,1)$ & $1$ [LD$$] & $(1,\sqrt{d})$ & $\sqrt{d}$  \\
       \hline     
                \hline
   Singularity  & \xmark  & \multicolumn{3}{c|}{good}   & bad \\  
    \hline 
    Brane-singularity conf. distance & \xmark  & \multicolumn{2}{c|}{$\infty$}   & \multicolumn{2}{c|}{finite} \\  
    \hline
    Timelike boundary   & \multicolumn{3}{c|}{\xmark}     & \multicolumn{2}{c|}{\cmark}    \\    
                \hline
    Bulk Black hole    & \multicolumn{4}{c|}{\cmark}  & \xmark \\
                \hline
    Gravity spectrum  $(D\geq 4)$   & \multicolumn{2}{c|}{continuum}  & gapped cont. & \multicolumn{2}{c|}{discretum} \\        
         \hline
         \hline
          Confinement & \multicolumn{2}{c|}{\xmark }  &     \multicolumn{3}{c|}{\cmark }   \\ 
                   \hline         
        Thermal phase transition $(D=5)$ & \multicolumn{3}{c|}{  second order }    & \multicolumn{2}{c|}{  first order }   \\ 
         \hline   
        Entanglement phase transition & \multicolumn{2}{c|}{\xmark }  & second order   & \multicolumn{2}{c|} {first order }   \\ 
         \hline   
    \end{tabular} 
    \caption{\it Properties of the $\Mnu^-$ spacetime. The first block reviews global features, the second block  summarizes key holographic properties. 
    }
    \label{tab:summary}
\end{table}

\subsubsection*{Confinement}
We first revisit the notion of holographic confinement based on classical strings. In our  background the behavior of the string is straightforwardly understood.
We show that for $\nu<1$ ($\nu>1$) the string bends towards small (large) $r$, while it is straight for $\nu=1$. An equivalent coordinate-free statement is that the string living in the $\Mnu$ background 
is repelled from the string frame curvature singularity.

We further show that, when attached to the boundary/brane on which the dual theory is defined, the string is forced to have a straight profile for $\nu>1$. The resulting picture is that the theory is not confining for $\nu<1$, while it is perfectly confining for both $\nu=1$ (linear dilaton) and $\nu>1$, in each case with {the} same string tension.
We show that this picture provides an intuitive understanding of holographic confinement in asymptotically-AdS backgrounds.

\subsubsection*{Entanglement Entropy}

Having a solid understanding of the confinement criterion, we  turn to holographic entanglement entropy. 
We consider a  $d-1$ strip  localized on the brane at given time.  Two kinds of surfaces anchored to the boundary of the strip are possible: a smooth surface  and a square one that reaches the singularity.

Working first with a general warped metric, we show that a metric zero at finite conformal distance implies that  the square surface has finite area and ends on the curvature singularity.
 We point out that a necessary condition for HEE phase transition to occur is that the width of the smooth surface's base be bounded from above. We determine a sufficient condition for this to occur, that relies on the existence of  a singularity at finite conformal distance.  
 We  show that for such a singularity,  there exist necessarily two smooth surfaces of different areas anchored to a given strip. This  implies that the HEE phase transition is first order in this case.

These phenomena are explicitly shown to occur in the  $\MnuM$ background.
The base of the smooth surface is bounded from above for $\nu\geq1$, and for $\nu>1$ two different smooth surfaces exist. This confirms that 
 the doubling phenomenon is tied to the existence of a singularity at finite conformal distance. 
We derive approximate formulas for the areas of  the smooth surfaces in the generic dilaton gravity background. We also find an exact  formula  for the area in the linear dilaton background.

For $\nu<1$ the smooth surface has  minimal area for any width of the strip. 
In this case the entanglement entropy reaches a plateau at large strip width for $ \nu \geq 0$ if $d>2$ and for $\nu>0$ if $d=2$. 

For $\nu=1$, the entanglement entropy   features a transition between the smooth surface and the square surface  at finite strip width. From 
our exact formula for the smooth surface in the linear dilaton background, 
we find that a discontinuity occurs at second order.  Hence the linear dilaton stands out once again as a special,  critical, case. 

For $\nu>1$, there are two smooth surfaces and the square one. A transition between the  smooth surface of smaller area and the square surface occurs at finite strip width. The transition turns out to be of first order. {The behavior of the  areas and of the transition closely resembles that obtained from stringy setups in~\cite{Klebanov:2007ws,Jokela:2020wgs,Jokela:2023lvr,Fatemiabhari:2024aua}.} In  our setup  we find that this characteristic behavior is tied to the curvature singularity being at finite conformal distance.

\subsubsection*{Thermodynamics and Stability}

We compute the on-shell effective action of the system with no black hole (i.e.~at zero temperature) as a function of the classical value of the dilaton field and of the brane position. Upon gauge fixing of spacetime fluctuations, the scalar fluctuations reduce to a single degree of freedom, the radion. We derive the radion profile and compute the full quadratic action of the radion, which completes and cross checks the results from \cite{Fichet:2023dju}. Under a certain condition on the brane-localized potential, the radion is massive, which ensures stability of the system.

The existence of the planar black hole amounts, from the viewpoint of the brane, to the existence of a perfect fluid.  We study the thermodynamics of this fluid, including phase transitions,  extending the results from \cite{Fichet:2023dju}.

We show that for $\nu\leq 1$ the  fluid/black hole appears via a second order phase transition whenever the temperature is nonzero. In contrast, for $\nu>1$, the black hole appears above a nonzero critical temperature via a first order phase transition. 
In the linear dilaton case $\nu=1$, the black hole has always the same nonzero temperature, but appears via a second order phase transition. 

Remarkably, the order of these thermal {phase} transitions matches the one of the entanglement entropy phase transitions. 

In the black hole phase considered, the black hole is always thermodynamically stable in the sense of having positive heat capacity. 
The black hole temperature increases with its size for $\nu<1$, while it decreses with its size for $\nu>1$, which causes the latent heat to be negative. 
The black hole temperature admits an upper bound for $\nu<1$, and a lower bound for $\nu>1$, which is tied to the brane not being engulfed by the horizon. 

We demonstrate that, in the extreme situation where  the horizon approaches the brane, a dramatic instability  in the dilaton vev occurs, that makes the classical effective potential unbounded from below.

\begin{acknowledgments}

We thank Matti J\"arvinen, Javier Subils, Miguel \'Angel V\'azquez-Mozo and collaborators for their valuable correspondence, and Lucas de Souza for insightful comments regarding
App.~\ref{app:aprimelimit}.  The work of EM is supported by the
project PID2020-114767GB-I00 and by the Ram\'on y Cajal Program under
Grant RYC-2016-20678 funded by MCIN/AEI/10.13039/50\-1100011033 and by
``FSE Investing in your future'', by Junta de Andaluc\'{\i}a under
Grant FQM-225, and by the ``Pr\'orrogas de Contratos Ram\'on y Cajal''
Program of the University of Granada. The work of MQ is supported by the grant
PID2023-146686NB-C31 funded by MICIU/AEI/10.13039/501100011033/ and by
FEDER, EU. IFAE is partially funded by the CERCA program of the
Generalitat de Catalunya.
  
\end{acknowledgments}

\appendix 

\section{Derivation of Property\,\eqref{eq:aprimelimit}}
\label{app:aprimelimit}

We prove Prop.\,\eqref{eq:aprimelimit}.  The essence of the derivation is to show that, using the mean value theorem, any interval in the vicinity of the metric zero  contains a point  such that {$a^\prime(z)/a(z)$} is singular, no matter how small the interval is. {We make the hypothesis that {$a^\prime(z)/a(z)$} is \textit{strictly  monotonic} near the singularity. }

The mean value theorem states that for any function $a(z)$ continuous on $[z_1,z_2]$ and  differentiable on $(z_1,z_2)$
 there is at least one  point $z_{12}$  in $(z_1,z_2)$ such that 
\be
\log a(z_2)-\log a(z_1)= \frac{a'(z_{12})}{a(z_{12})} (z_2-z_1)\,.
\label{eq:meanvalue}
\ee
We apply the theorem in the vicinity of the metric zero,  $a(z_s)=0$. Following the convention from section \ref{se:PT_boundary}, spacetime is defined to the right of $z_s$, $z\geq z_s$. 
We place the small {closed}  interval to the right of the metric zero, and take the limit  $z_1\to z_s^+$  in \eqref{eq:meanvalue}.

To ensure that  $z_{12}$ converges when taking the limit  $z_1\to z_s^+$,  we implement this limit using a discrete sequence as follows. 
Define $z_1\equiv z_s+\frac{1}{n}$,  where $n\in \mathbb{N}$ starts at a value larger than $\frac{1}{z_2-z_s}$ and is taken to infinity. For each $n$ there is a value of $z_{12}$, denoted  $z_{12}\equiv z(n,z_2) = z_{n}$. Since $z_n\in (z_s,z_2)$, $(z_n)$ is a bounded sequence. Thus the Bolzano-Weierstrass theorem ensures that  there exists a subsequence $( z_{n_j})\equiv (\hat z_{j})$ of $(z_{n})$ that is convergent, i.e.~$\underset{{j\to \infty}}{\lim} \hat z_{j} =  \hat z$, with $\hat z=  \hat z(z_2)$.

Starting from \eqref{eq:meanvalue}, we have
\be
 \underset{{j\to\infty }}{\lim}\frac{a'(\hat z_{j})}{a({\hat z_{j})}} =
  \underset{{z \to \hat z }}{\lim} \frac{a'( z)}{a({ z)}} =  
 \underset{{j\to \infty}}{\lim} \frac{\log a(z_2)-\log a(z_{s}+n_j^{-1})}{z_2-z_{s}-n_j^{-1}}=\infty \,, \label{eq:limitaprime}
\ee
since $ \log(a(z))$ diverges at $z_s$ while  by assumption $z_s$ is finite.   That is, there is at least one point $\hat z$ in $(z_s,z_2)$ for which {$a^\prime(z)/a(z)$} is infinite.
The assumption that {$a^\prime(z)/a(z)$} is {strictly  monotonic} on the interval 
$(z_s,z_2)$ implies that $\log(a(z))$ is strictly concave or convex, which ensures that the intermediate point $\hat z$ is unique.

We can then approach $z_2$  to $z_s^+$, such that the intermediate point $\hat z(z_2)$ is forced to tend to $z_s^+$. Taking this limit in \eqref{eq:limitaprime},  we obtain 
\be
 \underset{{z_2\to z_s^+}}{\lim}\frac{a'(\hat z(z_2))}{a(\hat z(z_2))} = \underset{{\hat z\to z_s^+}}{\lim}\frac{a'(\hat z)}{a(\hat z)} =\infty \,. 
\ee
This proves Prop.\,\eqref{eq:aprimelimit}.

\section{Minimal Surfaces and Geodesics
\label{app:app_surface}}

We demonstrate that the equations for the smooth minimal surfaces  anchored to the strip in $\mathcal{M}_{\nu}$  can be mapped onto geodesic equations in $\mathcal{M}_{\bar{\nu}}$. From \eqref{eq: area_smooth} and \eqref{eq: f_equation}, substituting the metric factors in Einstein frame \eqref{eq:ds2_Mnu},  we obtain
\begin{equation}
    \frac{d \mathrm{Area}(\gamma_{\cup})}{dr} = \frac{2 A_{d-2}}{\eta r_b^{\nu^2} L^{d-2}}\frac{r^{2(d-2)+\nu^2}}{\sqrt{r^{2(d-1)}-r_0^{2(d-1)}}}\,,\;\;\;\;\;\frac{d u}{dr} = \frac{r_0^{d-1} L}{\eta r_b^{\nu^2}}\frac{r^{\nu^2-2}}{\sqrt{r^{2(d-1)}-r_0^{2(d-1)}}}\,.
\label{eq:area_equations}
\end{equation}
These expressions closely resemble the differential equations for geodesics in $\mathcal{M}_\nu$ spacetime:
\begin{equation}
    \frac{ds}{dr} = \frac{2}{\eta r_b^{\nu^2}}\frac{r^{\nu^2}}{\sqrt{r^2 - r_0^2}}\,,\;\;\;\;\;\;\;\;\frac{d x}{dr} = \frac{r_0 L}{\eta r_b^{\nu^2}}\frac{r^{\nu^2-2}}{\sqrt{r^{2}-r_0^{2}}}\,,
\label{eq:area_equations_1}
\end{equation}
being $s$ the geodesic length. 
By introducing the change of variable $y = r^{d-1}$, we explicitly relate the two sets of equations. Under this transformation, the minimal surface equations \eqref{eq:area_equations} are related to the geodesic ones \eqref{eq:area_equations_1} by
\begin{equation}
    \frac{d \mathrm{Area}(\gamma_{\cup})}{dy} = \frac{A_{d-2}}{d-1}\left(\frac{r_b}{L} \right)^{d-2}\frac{ds}{dy}\bigg|_{\nu\to\bar{\nu}}\,,\;\;\;\;\;\frac{d u}{dy} = \frac{r_b^{d-2}}{d-1}\frac{dx}{dy}\bigg|_{\nu\to\bar{\nu}}\,,
\label{eq:area_equations_2}
\end{equation}
where
\begin{equation}
    \bar{\nu}^2 = 1 + \frac{\nu^2-1}{d-1}\,.
\end{equation}
For example, the minimal surfaces anchored to the strip in AdS$_{d+1}$ are related to the geodesics of ${\cal M}_{\bar \nu} $ by $\bar \nu^2= \frac{d-2}{d-1}$.

\bibliographystyle{JHEP}
\normalem
\bibliography{biblio}

\end{document}